\documentclass[a4paper,UKenglish,cleveref, autoref, thm-restate]{lipics-v2021}
\title{Metric Dimension and Geodetic Set Parameterized by Vertex Cover\thanks{An extended abstract of this paper appeared in the proceedings of STACS~2025.}} %TODO Please add

\titlerunning{Metric Dimension and Geodetic Set Parameterized by Vertex Cover} %TODO optional, please use if title is longer than one line

\author{Florent Foucaud}{Université Clermont Auvergne, CNRS, Mines Saint-Étienne, Clermont Auvergne INP, LIMOS, 63000 Clermont-Ferrand, France\and \url{https://perso.limos.fr/ffoucaud}}{florent.foucaud@uca.fr}{https://orcid.org/0000-0001-8198-693X}{ANR project GRALMECO (ANR-21-CE48-0004), French government IDEX-ISITE initiative 16-IDEX-0001 (CAP 20-25), International Research Center ``Innovation Transportation and Production Systems'' of the I-SITE CAP 20-25.}

\author{Esther Galby}{Department of Computer Science and Engineering, Chalmers University of Technology and University of Gothenburg, Gothenburg, Sweden}{galby@chalmers.se}{https://orcid.org/0009-0004-5398-2770}{}

\author{Liana Khazaliya}{Technische Universit\"{a}t Wien, Vienna, Austria \and \url{https://www.ac.tuwien.ac.at/people/lkhazaliya/}}{lkhazaliya@ac.tuwien.ac.at}{https://orcid.org/my-orcid?orcid=0009-0002-3012-7240}{Vienna Science and Technology Fund (WWTF) [10.47379/ICT22029]; Austrian Science Fund (FWF) [10.55776/Y1329]; European Union's Horizon 2020 COFUND programme [LogiCS@TUWien, grant agreement No.\ 101034440].}

\author{Shaohua Li}{School of Computer Science and Engineering, Central South University, Changsha, China}{shaohua.li@csu.edu.cn}{https://orcid.org/0000-0001-8079-6405}{National Natural Science Foundation of China under Grant 62472449.}

\author{Fionn {Mc Inerney}}{Telef\'{o}nica Scientific Research, Barcelona, Spain \and \url{https://sites.google.com/view/fionn-mc-inerney/home?pli=1}}{fmcinern@gmail.com}{https://orcid.org/0000-0002-5634-9506}{Smart Networks and Services Joint Undertaking (SNS JU) under the EU's Horizon Europe and innovation programme under Grant Agreement No. 101139067 (ELASTIC).}

\author{Roohani Sharma}{University of Bergen, Bergen, Norway}{r.sharma@uib.no}{https://orcid.org/0000-0003-2212-1359}{}

\author{Prafullkumar Tale}{Indian Institute of Science Education and Research Pune, Pune, India \and \url{https://pptale.github.io/}}{prafullkumar@iiserpune.ac.in}{https://orcid.org/0000-0001-9753-0523}{}

\authorrunning{F. Foucaud, E. Galby, L. Khazaliya, S. Li, F. {Mc Inerney}, R. Sharma, and P. Tale} %TODO mandatory. First: Use abbreviated first/middle names. Second (only in severe cases): Use first author plus 'et al.'

\Copyright{Esther Galby, Liana Khazaliya, Fionn {Mc Inerney}, Roohani Sharma, and Prafullkumar Tale} %TODO mandatory, please use full first names. LIPIcs license is "CC-BY";  http://creativecommons.org/licenses/by/3.0/


 %TODO mandatory; please add comma-separated list of keywords

\nolinenumbers %uncomment to disable line numbering

\hideLIPIcs

\usepackage{complexity}
\usepackage{mathtools}
\usepackage{bm}
\usepackage{xspace}
\usepackage{mathdots}
\usepackage{float}
\usepackage{needspace}
\usepackage[framemethod=tikz]{mdframed}

\newcommand{\mdfull}{\textsc{Metric Dimension}\xspace}

\newcommand{\gsfull}{\textsc{Geodetic Set}\xspace}

\newcommand{\OO}{\mathcal{O}}

\newcommand{\calC}{\mathcal{C}}

\newcommand{\calO}{\mathcal{O}}

\newcommand{\true}{\texttt{True}}
\newcommand{\false}{\texttt{False}}
\newcommand{\bit}{\textsf{bin}}

\newcommand{\bitrep}{\textsf{bit-rep}}

\newcommand{\bitrepnullifier}{\textsf{nullifier}}
\newcommand{\bits}{\textsf{bits}}

\newcommand{\vc}{\mathtt{vc}}
\newcommand{\fvs}{\mathtt{fvs}}
\newcommand{\td}{\mathtt{td}}
\newcommand{\tw}{\mathtt{tw}}
\newcommand{\pw}{\mathtt{pw}}

\newcommand{\yes}{\textsc{Yes}}
\newcommand{\no}{\textsc{No}}
\newcommand{\PSAT}{\textsc{3-Partitioned-3-SAT}\xspace}
\newcommand{\ETH}{\textsf{ETH}}

\newtheoremstyle{noparentheses}
    {}{}{\itshape}{}%
    {\bfseries}{.}{ }%
    {\thmname{#1}\thmnumber{ #2}\thmnote{ {\mdseries #3}}}
\theoremstyle{noparentheses} 
\newtheorem*{theoremnp*}{Theorem}

\usepackage{color}

\newcommand{\defproblem}[3]{
\noindent\fbox{
\vspace{1mm}
  \begin{minipage}{0.96\textwidth}
  \begin{tabular*}{\textwidth}{@{\extracolsep{\fill}}lr} #1 \\ \end{tabular*}
  {\bf{Input:}} #2  \\
  {\bf{Question:}} #3
  \end{minipage}
  }
}

\usepackage[colorinlistoftodos,textsize=tiny]{todonotes}
\usepackage{tikz}
\usetikzlibrary{positioning,backgrounds,patterns,calc}

\tikzset{
  circ/.style = {circle,draw,fill,inner sep=1.3pt}
}

\newtheorem{reduction rule}{Reduction Rule}
\EventEditors{John Q. Open and Joan R. Access}
\EventNoEds{2}
\EventLongTitle{42nd Conference on Very Important Topics (CVIT 2016)}
\EventShortTitle{CVIT 2016}
\EventAcronym{CVIT}
\EventYear{2016}
\EventDate{December 24--27, 2016}
\EventLocation{Little Whinging, United Kingdom}
\EventLogo{}
\SeriesVolume{42}
\ArticleNo{23}
\begin{document}

\maketitle

%TODO mandatory: add short abstract of the document
\begin{abstract}
  For a graph $G$, a subset $S \subseteq V(G)$ is called a
\emph{resolving set} of $G$ if, for any two vertices $u,v \in V(G)$,
there exists a vertex $w \in S$ such that $d(w,u) \neq d(w,v)$.  The
\mdfull problem takes as input a graph $G$ on $n$
vertices and a positive integer $k$, and asks whether there exists a
resolving set of size at most $k$. In another \emph{metric-based graph
problem}, \gsfull, the input is a graph $G$ and an integer
$k$, and the objective is to determine whether there exists a subset
$S \subseteq V(G)$ of size at most $k$ such that, for any vertex
$u \in V(G)$, there are two vertices $s_1, s_2 \in S$ such that $u$ lies on a
shortest path from $s_1$ to $s_2$.

These two classical problems turn out to be intractable with 
respect to the natural parameter, i.e., the solution size, as well as
most structural parameters, including the feedback vertex set number
and pathwidth. Some of the very few existing tractable results state that they
are both \FPT\ with respect to the vertex cover number $\vc$.

More precisely, we observe that both problems admit
an \FPT\ algorithm running in
$2^{\mathcal{O}(\vc^2)} \cdot n^{\OO(1)}$ time, and
a kernelization algorithm that 
outputs a kernel with $2^{\OO(\vc)}$ vertices.
We prove that unless the Exponential Time Hypothesis (\ETH) fails,
\mdfull and \gsfull, even on graphs of bounded diameter,
do not admit
\begin{itemize}
\item an \FPT\ algorithm running in
$2^{o(\vc^2)} \cdot n^{\OO(1)}$ time, nor
\item a kernelization algorithm that does not increase the solution size and 
outputs a kernel with $2^{o(\vc)}$ vertices.
\end{itemize}
The versatility of our technique enables us to apply it to both these problems.

We only know of one other problem in the literature that admits such
a tight lower bound.
Similarly, the list of known problems with exponential lower bounds on the number of {\em vertices} in kernelized instances is very short.
\end{abstract}

% !TEX root = ./main.tex 
\section{Introduction}
\label{sec:intro}
In this article, we study two \emph{metric-based} graph problems,
one of which is defined through distances, while the other relies on shortest paths.
%These problems are defined using either distance values or shortest paths in the graph. 
Metric-based graph problems are
ubiquitous in computer science; for example, the classical
\textsc{(Single-Source) Shortest Path}, \textsc{(Graphic) Traveling
 Salesperson} or \textsc{Steiner Tree} problems fall into this
category. 
Those are fundamental problems, often stemming from
applications in network design, for which a considerable amount of algorithmic research
has been done. 
%Among these, 
Metric-based graph packing and
covering problems, like \textsc{Distance
 Domination}~\cite{JKST19} or \textsc{Scattered Set}~\cite{KLP22},
have recently gained a lot of attention. 
Their non-local nature leads to non-trivial algorithmic properties 
that differ from most
graph problems with a more local nature. 
%This is the case in particular for treewidth-based algorithms. 
%In this paper,

We focus here on the \mdfull and \gsfull problems, which arise from
network monitoring and network design, respectively. As noted in the
introduction of~\cite{floISAAC20} and the conclusion of~\cite{KK22},
and recently demonstrated in~\cite{BDM23,DBLP:journals/corr/abs-2307-08149},
these two metric-based graph covering problems share many algorithmic
properties.
%% This is confirmed in
%% the present paper, where we show that these two problems have the same
%% properties with respect to the vertex cover parameter.
They have far-reaching applications, as exemplified by, 
e.g., the recent work~\cite{BDM23} where
it was shown that enumerating minimal solution sets for 
\mdfull and \gsfull in (general) graphs and split graphs, respectively, is equivalent to 
the enumeration of minimal transversals in hypergraphs, whose solvability in total-polynomial time is arguably
the most important open problem in algorithmic enumeration.
%We formally define these two problems before proceeding.
Formally, these two problems are defined as follows.

\smallskip
\defproblem{\mdfull}{A graph $G$ on $n$ vertices and a positive integer $k$.}{Does there exist $S \subseteq V(G)$ such that $|S| \leq k$ and, for any pair of vertices $u,v\in V(G)$,
there exists a vertex $w\in S$ with $d(w,u)\neq d(w,v)$?}

\smallskip
\defproblem{\gsfull}{A graph $G$ on $n$ vertices and a positive integer $k$.}{Does there exist $S \subseteq V(G)$ such that $|S| \leq k$ and, for any vertex $u \in V(G)$, there are two
vertices $s_1, s_2 \in S$ such that $u$ lies on a shortest path from $s_1$ to
$s_2$?}
\smallskip

\mdfull dates back to the 1970s~\cite{HM76,Slater75},
whereas \gsfull was introduced in 1993~\cite{harary1993}. 
The non-local nature of these problems 
has since posed interesting algorithmic challenges.
\mdfull was first shown to be \NP-complete in general graphs in Garey and
Johnson's book~\cite{GJ79}, and this was later extended to many
restricted graph classes (see `Related work' below). 
\gsfull was proven to be \NP-complete in the seminal paper~\cite{harary1993},
and later shown to be \NP-hard on restricted graph 
classes as well.

As these two problems are \NP-hard even in very restricted cases,
it is natural to ask for ways to confront this hardness.
In this direction, the parameterized complexity paradigm allows for a more refined analysis of a problem’s complexity.
In this setting, we associate each instance $I$ of a problem with a parameter $\ell$, and are interested in algorithms running in $f(\ell) \cdot |I|^{\OO(1)}$ time for some computable function $f$.
Parameterized problems that admit such an algorithm are called fixed-parameter tractable (\FPT\ for short) with respect to the considered parameter.
Under standard complexity assumptions, parameterized problems that are hard for the complexity class \W[1] or \W[2] do not admit such algorithms.
A parameter may originate from the formulation of the problem itself (called a \emph{natural parameter}) or it can be a property of the input (called a \emph{structural parameter}).

This approach, however, had limited success in the case
of these two problems.
In the seminal paper~\cite{HartungN13}, \mdfull was proven to be
\W[2]-hard parameterized by the solution size~$k$, even in subcubic
bipartite graphs. 
Similarly, \gsfull is \W[2]-hard parameterized by the solution size~\cite{dourado2010,KK22}, even on chordal bipartite graphs.
%Similarly, \gsfull is known to be \W[1]-hard parameterized by the solution size~\cite{KK22}.
These initial hardness results drove the ensuing meticulous study of the problems under structural
parameterizations.
We present an overview of such results in `Related work' below. 
In this article, we focus on the \emph{vertex cover number},
denoted by $\vc$, of the input graph and prove 
the following positive results.

\begin{theorem}
\label{thm:algo-vertex-cover}
\mdfull and \gsfull admit
\begin{itemize}
\item \FPT\ algorithms running in
$2^{\mathcal{O}(\vc^2)} \cdot n^{\OO(1)}$ time, and
\item kernelization algorithms that 
output kernels with $2^{\OO(\vc)}$ vertices.
\end{itemize}
\end{theorem}

The second set of results follows from simple reduction rules, and was
also observed in~\cite{HartungN13} for \mdfull.
The first set of results builds on the second set by using a simple,
but critical observation.
For \mdfull, this also improves upon the $2^{2^{\mathcal{O}(\vc)}} \cdot n^{\OO(1)}$
algorithm mentioned in~\cite{HartungN13}.
Our main technical contribution, however, is in proving that
these results are optimal assuming the Exponential Time Hypothesis (\ETH).

\begin{theorem}
\label{thm:lower-bound-vertex-cover}
Unless the \ETH\ fails,
\mdfull and \gsfull do not admit
\begin{itemize}
\item \FPT\ algorithms running in
$2^{o(\vc^2)} \cdot n^{\OO(1)}$ time, nor
\item kernelization algorithms that do not increase the solution size and 
output kernels with $2^{o(\vc)}$ vertices,
\end{itemize}
even on graphs of bounded diameter.
\end{theorem}

Both of these statements constitute a rare set of results in the existing literature.
We know of only one other problem in the literature that admits a lower
bound of the form $2^{o(\vc^2)} \cdot n^{\OO(1)}$ and a matching upper
bound~\cite{DBLP:journals/toct/AgrawalLSZ19} - whereas such results
parameterized by pathwidth are mentioned in~\cite{DBLP:conf/mfcs/Pilipczuk11} and~\cite{DBLP:journals/iandc/SauS21}. 
Very recently, the authors in~\cite{DBLP:journals/corr/abs-2402-08346}
also proved a similar result with respect to solution size.
Similarly, the list of known problems with exponential lower bounds on the number of {\em vertices} in kernelized instances is very short.\footnote{For the definition of a kernelized instance and kernelization algorithm, refer to Section~\ref{sec:preliminaries} or
  \cite{DBLP:books/sp/CyganFKLMPPS15}.} 
To the best of our knowledge, the only known results of this kind (that is, \ETH-based lower bounds on the number of vertices in a kernel) are for
\textsc{Edge Clique Cover}~\cite{DBLP:journals/siamcomp/CyganPP16},
\textsc{Biclique Cover}~\cite{DBLP:conf/iwpec/ChandranIK16},
\textsc{Steiner Tree}~\cite{MPP18},
\textsc{Strong Metric Dimension}~\cite{DBLP:journals/corr/abs-2307-08149},
\textsc{B-NCTD$^+$}~\cite{CCMR23}, \textsc{Locating Dominating Set}~\cite{DBLP:journals/corr/abs-2402-08346}, and \textsc{Telephone Broadcasting}~\cite{DBLP:journals/corr/abs-2403-03501}.\footnote{{\sc Point Line Cover} also does not admit a kernel with $\OO(k^{2-\epsilon})$ {\em vertices}, for any $\epsilon >0$, unless $\NP \subseteq \coNP/poly$~\cite{DBLP:journals/talg/KratschPR16}.}
For \mdfull, the above also improves a result
of~\cite{DBLP:journals/tcs/GutinRRW20}, which states that \mdfull\ parameterized by $k+\vc$ does not admit a polynomial kernel unless the polynomial hierarchy collapses to its third level. Indeed, the result
of~\cite{DBLP:journals/tcs/GutinRRW20} does not rule out a kernel of super-polynomial or sub-exponential size.

In a recent work~\cite{DBLP:journals/corr/abs-2307-08149}, the present set of authors proved that, 
unless the \ETH\ fails,
\mdfull and \gsfull on graphs of bounded diameter do not admit
$2^{2^{o(\tw)}} \cdot n^{\OO(1)}$-time algorithms,
thereby establishing one of the first such results for
\NP-\emph{complete} problems. 
Note that $n \succ \vc \succ \fvs \succ \tw$ and $n \succ \vc
\succ \td \succ \pw \succ \tw$ in the parameter hierarchy, where $n$ is the number of vertices, $\fvs$ 
is the feedback vertex set number, $\td$ is the treedepth,
and $\tw$ is the treewidth of the graph.
The authors further proved that their lower bound
also holds for $\fvs$ and $\td$ in the case of \mdfull, and for $\td$ in the case of \gsfull~\cite{DBLP:journals/corr/abs-2307-08149}.
Note that a simple brute-force algorithm enumerating all possible candidates
runs in $2^{\calO(n)}$ time for both of these problems.
%Hence, there is a natural gap of single-exponential algorithm and
%double-exponential 
Thus, the next natural question is whether such a lower bound for 
\mdfull and \gsfull\ can be extended to larger
parameters, in particular $\vc$.
Our first results answer this question in the negative.
Together with the lower bounds with respect to $\vc$, 
this establishes the boundary between parameters yielding 
single-exponential and double-exponential running
times for \mdfull\ and \gsfull.

Before moving forward, we highlight the parallels and differences
between Foucaud et al.~\cite{DBLP:journals/corr/abs-2307-08149} 
and our work. 
Their aim was to establish double-exponential
lower bounds for \NP-complete problems, and to do so they focused
on the restriction of the problems to graphs of bounded treewidth and diameter. 
Our objective is to closely examine one of the very few 
tractable results for \mdfull\ and \gsfull\ on general graphs by focusing
on the vertex cover parameter.
While we use some gadgets from~\cite{DBLP:journals/corr/abs-2307-08149}, overall our reductions significantly differ from the corresponding reductions in that article.
Note that we need to ``control''  the vertex cover number
of the reduced graph, whereas the corresponding 
reductions by  Foucaud et al.~\cite{DBLP:journals/corr/abs-2307-08149}
only need to ``control'' the treewidth.

% !TEX root = ./main.tex
%\section{Related Work}\label{sec:related-work}

%\subsection{Metric Graph Problems}\label{sec:related-work-metric}

\subparagraph*{Related Work.}
We mention here results concerning structural parameterizations 
of \mdfull and \gsfull, and refer the reader to the full version of~\cite{DBLP:journals/corr/abs-2307-08149}
for a more comprehensive overview of applications and
related work regarding these two problems.

As previously mentioned, \mdfull is
\W[2]-hard parameterized by the solution size $k$, even in subcubic
bipartite graphs~\cite{HartungN13}. 
%This seminal paper motivated the subsequent meticulous study of \mdfull under structural parameterizations. 
Several other parameterizations have 
been studied for this problem, on which we elaborate next (see also~\cite[Figure~1]{GKIST23}).
Through careful algorithmic design, kernelization, and/or meta-theorems, 
it was proven that there is an \XP\ algorithm parameterized by the 
feedback edge set number \cite{ELW15}, and \FPT\ algorithms 
parameterized by the max leaf number \cite{E15}, the modular-width 
and the treelength plus the maximum degree \cite{BelmonteFGR17}, 
the treedepth and the clique-width plus the diameter \cite{GHK22}, 
and the distance to cluster (co-cluster, respectively) \cite{GKIST23}. 
Recently, an \FPT\ algorithm parameterized by the treewidth in chordal 
graphs was given in~\cite{BDP23}. 
On the negative side, \mdfull is \W[1]-hard parameterized by the pathwidth even 
on graphs of constant degree~\cite{BP21}, para-\NP-hard parameterized 
by the pathwidth~\cite{LM21}, and \W[1]-hard parameterized by 
the combined parameter feedback vertex set number plus pathwidth~\cite{GKIST23}. 

The parameterized complexity of \gsfull was first addressed in~\cite{KK22}, in which they observed that the reduction
from~\cite{dourado2010} implies that the problem is \W[2]-hard 
parameterized by the solution size (even for chordal
bipartite graphs). This motivated the authors of~\cite{KK22} to investigate
structural parameterizations of \gsfull. They proved the problem to
be \W[1]-hard for the combined parameters solution size, feedback vertex set
number, and pathwidth, and \FPT\ for the parameters treedepth,
modular-width (more generally, clique-width plus diameter), and
feedback edge set number~\cite{KK22}. The problem was also shown to be \FPT\ on
chordal graphs when parameterized by the treewidth~\cite{floISAAC20}.

% !TEX root = ./main.tex
\section{Preliminaries}
\label{sec:preliminaries}

For an integer $a$, we let $[a] = \{1,\ldots,a\}$.

\subparagraph*{Graph theory.} 
We use standard graph-theoretic notation and refer the reader 
to~\cite{D12} for any undefined notation. 
For an undirected graph $G$, the sets $V(G)$ and $E(G)$ denote its 
set of vertices and edges, respectively.
Two vertices $u,v\in V(G)$ are {\it adjacent} or {\it neighbors} if 
$(u, v)\in E(G)$. 
The {\it open neighborhood} of a vertex $u\in V(G)$, denoted by $N(u):=N_G(u)$, is the set of vertices that are neighbors of $u$. 
The {\it closed neighborhood} of a vertex $u\in V(G)$ is denoted by $N[u]:=N_G[u]:=N_G(u)\cup \{u\}$.
For any $X \subseteq V(G)$ and $u\in V(G)$, $N_X(u) = N_G(u) \cap X$. 
Any two vertices $u,v\in V(G)$ are {\it true twins} if $N[u] = N[v]$, and are {\it false twins} if $N(u) = N(v)$. 
Observe that if $u$ and $v$ are true twins, then $(u,v) \in E(G)$, but if they are only false twins, then $(u,v) \not \in E(G)$.
For a subset $S$ of $V(G)$, we say that the vertices in $S$ are true (false, respectively) twins if, for any $u,v\in S$, $u$ and $v$ are true (false, respectively) twins.
The {\it distance} between two vertices $u,v\in V(G)$ in $G$, denoted by $d(u,v):=d_G(u,v)$, is the length of a $(u,v)$-shortest path in $G$. 
For a subset $S$ of $V(G)$, we define $N[S] = \bigcup_{v \in S} N[v]$ and $N(S) = N[S] \setminus S$.
%For a subset $S$ of $V(G)$, we denote the graph obtained by deleting $S$ from $G$ by $G - S$.
For a graph $G$, a set $X \subseteq V(G)$ is said to be {a} \emph{vertex cover} if $V(G) \setminus X$ is an independent set. We denote by $\vc(G)$  the size of a minimum vertex cover in $G$. When $G$ is clear from the context, we simply say $\vc$.
%For a graph $G$, a set $X \subseteq V(G)$ is said to be {a} \emph{feedback vertex set} of $G$ if $V(G) \setminus X$ is an acyclic graph. We define the notation of \emph{the feedback vertex set number} in the analogous way.

\subparagraph*{Metric Dimension and Geodetic Set.}
A subset of vertices $S\subseteq V(G)$ {\it resolves} a pair of vertices $u,v\in V(G)$ if there exists a vertex $w \in S$ such that $d(w,u)\neq d(w,v)$.
A subset of vertices $S\subseteq V(G)$ is a {\it resolving set} of $G$ if it resolves all pairs of vertices $u,v\in V(G)$.
A vertex $u\in V(G)$ is {\it distinguished} by a subset of vertices $S\subseteq V(G)$ if, for any $v\in V(G)\setminus \{u\}$, there exists a vertex $w\in S$ such that $d(w,u)\neq d(w,v)$.
%For an ordered subset of vertices $S=\{s_1,\dots,s_k\}\subseteq V(G)$ and a single vertex $u\in V(G)$, the {\it distance vector} of $S$ with respect to $u$ is $r(S|u):=(d(s_1,u),\dots,d(s_k,u))$.

\begin{observation}\label{obs:twins}
Let $G$ be a graph. 
For any (true or false) twins $u,v \in V(G)$ and any $w \in V(G) \setminus \{u,v\}$, $d(u,w) = d(v,w)$, 
and so, for any resolving set $S$ of $G$, $S\cap \{u,v\} \neq \emptyset$.
\end{observation}
\begin{proof}
As $w\in V(G) \setminus \{u,v\}$, and $u$ and $v$ are (true or false) twins, the shortest $(u,w)$- and $(v,w)$-paths contain a vertex of $N:=N(u)\setminus\{v\}=N(v)\setminus\{u\}$, and $d(u,w)=d(v,w)$. Hence, any resolving set $S$ of $G$ contains at least one of $u$ and $v$.
\end{proof}

A subset $S \subseteq V(G)$ is a \emph{geodetic set} if for every $u \in V(G)$, the following holds: there exist $s_1,s_2 \in S$ such that $u$ lies on a shortest path from $s_1$ to $s_2$.
The following simple observation is used throughout the paper. Recall that a vertex is \emph{simplicial} if its neighborhood forms a clique.
Observe that any simplicial vertex $v$ does not belong to any shortest path between any pair $x,y$ of vertices (both distinct from $v$).
Hence, the following observation follows:

\begin{observation}[\cite{CHZ02}]\label{obs:simplicial}
If a graph $G$ contains a simplicial vertex $v$, then $v$ belongs to any geodetic set of $G$.
Specifically, every degree-$1$ vertex belongs to any geodetic set of $G$.
\end{observation}

\subparagraph*{Parameterized Complexity.}
An instance of a parameterized problem $\Pi$ comprises an input $I$, which is an input of the classical instance of the problem, and an integer $\ell$, which is called the parameter.
A problem $\Pi$ is said to be \emph{fixed-parameter tractable} or in \FPT\ if given an instance $(I,\ell)$ of $\Pi$, we can decide whether or not $(I,\ell)$ is a \yes-instance of $\Pi$ in $f(\ell)\cdot |I|^{\OO(1)}$ time,
for some computable function $f$ whose value depends only on $\ell$. 

A {\em kernelization} algorithm for $\Pi$ is a polynomial-time algorithm that takes as input an instance $(I,\ell)$ of $\Pi$ and returns an {\em equivalent} instance $(I',\ell')$ of $\Pi$, where $|I'|, \ell' \leq f(\ell)$, where $f$ is a function that depends only on the initial parameter $\ell$. If such an algorithm exists for $\Pi$, we say that $\Pi$ admits a kernel of {\em size} $f(\ell)$. If $f$ is a polynomial or exponential function of $\ell$, we say that $\Pi$ admits a polynomial or exponential kernel, respectively. 
If $\Pi$ is a graph problem, then $I$ contains a graph, say $G$, and $I'$ contains a graph, say $G'$. In this case, we say that $\Pi$ admits a kernel with $f(\ell)$ vertices if the number of vertices of $G'$ is at most $f(\ell)$. 

It is typical to describe a kernelization algorithm as a series of reduction rules.
A \emph{reduction rule} is a polynomial-time algorithm that takes as an input an instance of a problem and outputs another (usually reduced) instance.
A reduction rule said to be \emph{applicable} on an instance if the output instance is different from the input instance.
A reduction rule is \emph{safe} if the input instance is a \yes-instance if and only if the output instance is a \yes-instance.

The Exponential Time Hypothesis roughly states that $n$-variable {\sc 3-SAT} cannot be solved in $2^{o(n)}$ time.
For more on parameterized complexity and related terminologies, we refer the reader to the recent book by Cygan et al.~\cite{DBLP:books/sp/CyganFKLMPPS15}.

\subparagraph*{\textsc{3-Partitioned-3-SAT}.}
Our lower bound proofs consist of reductions from the \textsc{3-Partitioned-3-SAT} problem.
This version of \textsc{3-SAT} was introduced
in~\cite{DBLP:journals/corr/abs-2302-09604} and is defined as follows.

\smallskip
\defproblem{\textsc{3-Partitioned-3-SAT}}{A formula $\psi$ in $3$-\textsc{CNF} form,
together with a partition of the set of its variables into three
disjoint sets $X^{\alpha}$, $X^{\beta}$, $X^{\gamma}$,
with $|X^{\alpha}| = |X^{\beta}| = |X^{\gamma}| = n$, and
such that no clause contains more than one variable from each of
$X^{\alpha},  X^{\beta}$, and $X^{\gamma}$.}{Determine whether $\psi$ is satisfiable.}
\smallskip

The authors of~\cite{DBLP:journals/corr/abs-2302-09604} also proved the following.

\begin{proposition}[{\cite[Theorem 3]{DBLP:journals/corr/abs-2302-09604}}]
\label{prop:3-SAT-to-3-Partition-3-SAT-lampis}
Unless the \ETH\ fails,  \textsc{3-Partitioned-3-SAT} does not admit
an algorithm running in $2^{o(n)}$ time.
\end{proposition}
% !TEX root = ./main.tex
\section{\mdfull: Lower Bounds Regarding Vertex Cover}
\label{sec:lower-bound-vertex-cover}

In this section, we first prove
the following theorem.

\begin{theorem}
\label{thm:reduction-3-SAT-Met-Dim-VC}
There is an algorithm that, given an instance
$\psi$ of \textsc{3-Partitioned-3-SAT} on~$N$ variables,  
runs in $2^{\calO(\sqrt{N})}$ time, and constructs an equivalent 
instance $(G, k)$ of \mdfull such that 
$\vc(G) + k = \OO(\sqrt{N})$ (and $|V(G)| = 2^{\OO(\sqrt{N})}$).
\end{theorem}

The above theorem, along with the 
arguments that are standard to prove 
the \ETH-based lower bounds, 
immediately imply the following 
results.

\begin{corollary}
\label{cor:MD-vc-lower-bound-algo}
Unless the \ETH\ fails, 
\mdfull does not admit an algorithm 
running in $2^{o(\vc^2)} \cdot n^{\OO(1)}$ time. 
\end{corollary}
\begin{corollary}
\label{cor:MD-vc-lower-bound-kernel}
Unless the \ETH\ fails, 
\mdfull does not admit a kernelization algorithm that
does not increase the solution size $k$ and
outputs a kernel with $2^{o(k + \vc)}$ vertices. 
\end{corollary}
\begin{proof}For the sake of contradiction, assume that such a kernelization algorithm exists.
Consider the following algorithm for \textsc{$3$-SAT}.
Given a \textsc{$3$-SAT} formula on $N$ variables, 
it uses Theorem~\ref{thm:reduction-3-SAT-Met-Dim-VC} to obtain
an equivalent instance of $(G, k)$ such that $\vc(G) + k = \calO(\sqrt{N})$ and $|V(G)| = 2^{\OO(\sqrt{N})}$.
Then,  it uses the assumed kernelization algorithm
to construct an equivalent instance $(H, k')$ such that 
$H$ has $2^{o(\vc(G) + k)}$ vertices and
$k' \le k$.
Finally,  it uses a brute-force algorithm, running in $|V(H)|^{\calO(k')}$ time,
to determine whether the reduced instance, or equivalently the input
instance of \textsc{$3$-SAT},  is a \yes-instance.
The correctness of the algorithm follows from the 
correctness of the respective algorithms and our assumption.
The total running time of the algorithm is 
$2^{\calO(\sqrt{N})} + (|V(G)| + k)^{\calO(1)} + |V(H)|^{\calO(k')}
= 2^{\calO(\sqrt{N})} + (2^{\calO(\sqrt{N})})^{\calO(1)} + (2^{o(\sqrt{N})})^{\calO(\sqrt{N})}
= 2^{o(N)}$.
But this contradicts the \ETH.
\end{proof}

Before presenting the reduction, we first introduce some preliminary tools.

%\subsection{Preliminary Tool: Set Representation}
\subsection{Preliminary Tools}
\label{subsec:prelim-3-Par-3-SAT-Met-Dim-diam-tw}

\begin{figure}[t]
    \centering
        \includegraphics[scale=0.93]{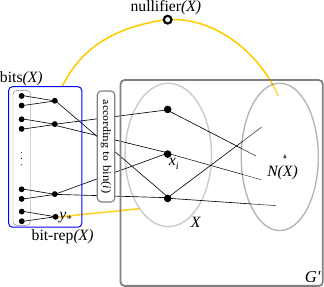} \hspace{0.4cm}
        \includegraphics[scale=0.93]{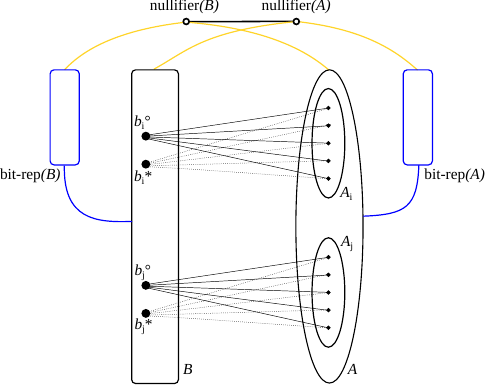}
    \caption{ \textbf{Set Identifying Gadget (left)}. The blue box represents $\bitrep(X)$. The yellow lines represent that all possible edges exist between $\bitrep(X)\setminus \bits(X)$ and $\bitrepnullifier(X)$, $\bitrepnullifier(X)$ and $N(X)$, and $y_{\star}$ and $X$. Note that $G'$ is not necessarily restricted to the graph induced by the vertices in $X\cup N(X)$.
    \textbf{Vertex Selector Gadget (right)}.
    For $X \in \{B,  A\}$,  the blue box represents $\bitrep(X)$, the blue link represents the connection
    with respect to the binary representation, and the yellow line represents that $\bitrepnullifier(X)$
    is adjacent to each vertex in $\bitrep(X)\setminus \bits(X)$.
    Dotted lines highlight absent edges.}
    \label{fig:set-identifying-gadget}
    \label{fig:vertex-selector-gadget}
\end{figure}

\subsubsection{Set Identifying Gadget}
\label{subsubsec:gadget-set-MD}

We redefine a gadget introduced in~\cite{DBLP:journals/corr/abs-2307-08149}. Suppose we are given a graph $G'$ and a subset $X\subseteq V(G')$ of its vertices.
Further, suppose that we want to add a vertex set $X^+$ to $G'$ in order to obtain a new graph $G$ such that (1) each vertex in $X \cup X^+$ will be distinguished by vertices in $X^+$ that must be in any resolving set $S$ of $G$, and (2) no vertex in $X^+$ can resolve any pair of vertices in $V(G)\setminus (X\cup X^+)$ that are in the same distance class with respect to~$X$.
The graph induced by the vertices of $X^+$, along with the edges connecting $X^+$ to $G'$, is referred to as the Set Identifying Gadget for the set $X$~\cite{DBLP:journals/corr/abs-2307-08149}.

Given a graph $G'$ and a non-empty subset $X\subseteq V(G')$ of its vertices, to construct such a graph $G$, we add vertices and edges to $G'$ as follows:
\begin{itemize}
\item The vertex set $X^+$ that we are aiming to add is the union of a set $\bitrep(X)$ and a special vertex denoted by $\bitrepnullifier(X)$.
\item Let $X=\{x_i\mid i\in [|X|]\}$ and set $q := \lceil \log(|X|+ 2) \rceil+1$.
We select this value for $q$ to (1)~uniquely represent each integer in $[|X|]$ by its bit-representation in binary (note that we start from $1$ and not $0$), (2)~ensure that the only vertex whose bit-representation contains all $1$'s is $\bitrepnullifier(X)$, and (3)~reserve one spot for an additional vertex $y_{\star}$.
\item For every $i \in [q]$, add three vertices $y^a_i,  y_i, y^b_i$, and add the path $(y^a_i, y_i, y^b_i)$.
\item Add three vertices $y^a_{\star},  y_{\star}, y^b_{\star}$, and add the path $(y^a_{\star}, y_{\star}, y^b_{\star})$. 
Add all the edges to make $\{y_i\mid\ i \in [q] \}\cup\{y_{\star}\}$ a clique.
Make $y_{\star}$ adjacent to each vertex $v\in X$.
We denote $\bitrep(X)=\{y_i, y^a_i, y^b_i\mid i\in [q]\}\cup \{y_{\star}, y^a_{\star}, y^b_{\star}\}$ and its subset $\bits(X)=\{y^a_i, y^b_i\mid i\in [q]\}\cup \{y^a_{\star}, y^b_{\star}\}$ for convenience in a later case analysis.
\item For every integer $j \in [|X|]$, let $\bit(j)$ denote the binary
representation of $j$ using $q$ bits.
Connect $x_j$ with $y_{i}$
if the $i^{th}$ bit (going from left to right) in $\bit(j)$ is $1$.
\item Add a vertex, denoted by $\bitrepnullifier(X)$,
and make it adjacent to every vertex in $\{y_i\mid i \in [q] \}\cup\{y_{\star}\}$.
One can think of $\bitrepnullifier(X)$ as the only vertex whose bit-representation contains all $1$'s.
\item For every vertex $u \in V(G)\setminus (X\cup X^+)$ such that
$u$ is adjacent to some vertex in $X$, add an edge between
$u$ and $\bitrepnullifier(X)$.
We add this vertex to ensure that
vertices in $\bitrep(X)$ do not resolve critical pairs
in $V(G)$.
\end{itemize}
This completes the construction of $G$. The properties of $G$ are not proven yet, but just given as an intuition behind its construction. See Figure~\ref{fig:set-identifying-gadget} for an illustration.

\subsubsection{Gadget to Add Critical Pairs}
\label{subsubsec:gadget-critical-pairs}

Any resolving set needs to resolve \emph{all} pairs of vertices in
the input graph.
As we will see, some pairs, which we call critical pairs, are harder
to resolve than others.
In fact, the non-trivial part will be to
resolve all of the critical pairs.

Suppose that we need to have $m \in \mathbb{N}$ critical pairs in a graph $G$, say $\langle c^\circ_i,   c^\star_i \rangle$ for every $i \in [m]$.
Define $C := \{c^{\circ}_i,  c^{\star}_i\mid i \in [m]\}$.
We then add $\bitrep(C)$ and $\bitrepnullifier(C)$ as mentioned above (taking $C$ as the set $X$), with
the edges between $\{c^{\circ}_i, c^\star_i\}$ and $\bitrep(C)$ defined by $\bit(i)$, i.e.,
connect both $c^{\circ}_i$ and $c^\star_i$ with the $j$-th vertex of $\bitrep(C)$
if the $j^{th}$ bit (going from left to right) in $\bit(i)$ is $1$.
Hence,  $\bitrep(C)$ can resolve any pair of the form
$\langle c^{\circ}_i,  c^{\star}_{\ell} \rangle$,  $\langle c^{\circ}_i,  c^{\circ}_{\ell} \rangle$,  or
$\langle c^{\star}_i,  c^{\star}_{\ell} \rangle$ as long as $i \neq \ell$.
As before, $\bitrep(C)$ can also resolve all pairs with one vertex in
$C \cup \bitrep(C) \cup \{\bitrepnullifier(C)\}$, but no critical pair of vertices. Again, when these facts
will be used, they will be proven formally.

\subsubsection{Vertex Selector Gadgets}
\label{subsubsec:vertex-selector}

Suppose that we are given a collection of sets $A_1, A_2, \dots, A_q$ of vertices
in a graph $G$, and
we want to ensure that any resolving set of $G$ includes at least one vertex
from $A_i$ for every $i \in [q]$.
In the following,  we construct a gadget that achieves a slightly weaker objective.
\begin{itemize}
\item Let $A = \underset{i \in [q]}{\bigcup} A_i$.
Add a set identifying gadget for $A$ as mentioned in Subsection~\ref{subsubsec:gadget-set-MD}.
\item For every $i \in [q]$,  add two vertices $b^{\circ}_i$ and $b^{\star}_i$.
Use the gadget mentioned in Subsection~\ref{subsubsec:gadget-critical-pairs}
to make all the pairs of the form
$\langle b^{\circ}_i, b^{\star}_i \rangle$ critical pairs.

\item For every $a \in A_i$,  add an edge $(a, b^{\circ}_i)$.
We highlight that we do not make $a$ adjacent to $b^{\star}_i$
by a dotted line in Figure~\ref{fig:vertex-selector-gadget}.
Also, add the edges $(a,  \bitrepnullifier(B))$,  $(b^{\circ}_i,  \bitrepnullifier(A))$, $(b^{\star}_i,  \bitrepnullifier(A))$, and $(\bitrepnullifier(A), \bitrepnullifier(B))$.
\end{itemize}
This completes the construction.

Note that the only vertices that can resolve a critical pair $\langle b^{\circ}_i, b^{\star}_i \rangle$,
apart from $b^{\circ}_i$ and $b^{\star}_i$, are the vertices in $A_i$.
Hence, every resolving set contains at least one vertex in
$\{b^{\circ}_i, b^{\star}_i\} \cup A_i$. Again, when used, these facts will be proven formally.

\subsection{Reduction}
\label{subsec:reduction-3-Par-3-SAT-Met-Dim}

%In this subsection,  we present a reduction from
%\textsc{3-Partitioned-3-SAT} to \textsc{Metric Dimension}.
%As mentioned before, this version of \textsc{3-SAT} was introduced
%in~\cite{DBLP:journals/corr/abs-2302-09604} and is defined as follows.
%
%\defproblem{\textsc{3-Partitioned-3-SAT}}{A formula $\psi$ in $3$-\textsc{CNF} form,
%together with a partition of the set of its variables into three
%disjoint sets $X^{\alpha}$, $X^{\beta}$, $X^{\gamma}$,
%with $|X^{\alpha}| = |X^{\beta}| = |X^{\gamma}| = n$, and
%such that no clause contains more than one variable from each of
%$X^{\alpha},  X^{\beta}$, and $X^{\gamma}$.}{Determine whether $\psi$ is satisfiable.}
%
%The authors of~\cite{DBLP:journals/corr/abs-2302-09604} also proved the following.
%
%\begin{proposition}[{\cite[Theorem 3]{DBLP:journals/corr/abs-2302-09604}}]
%\label{prop:3-SAT-to-3-Partition-3-SAT}
%Unless the \ETH\ fails,  \textsc{3-Partitioned-3-SAT} does not admit
%an algorithm running in $2^{o(n)}$ time.
%\end{proposition}

Consider an instance $\psi$ of \textsc{3-Partitioned-3-SAT}
with $X^{\alpha}, X^{\beta}, X^{\gamma}$ the partition of the variable set.
By adding dummy variables in each of these sets, we can assume that $\sqrt{n}$ is an integer.
From $\psi$, we construct the graph $G$ as follows.
We describe the construction of the part of the graph $G$ that corresponds to $X^{\alpha}$, with the parts corresponding to $X^{\beta}$ and $X^{\gamma}$ being analogous.
We rename the variables in $X^{\alpha}$
to $x^{\alpha}_{i, j}$ for $i, j \in [\sqrt{n}]$.

\begin{figure}[t]
    \centering
        \includegraphics[scale=1]{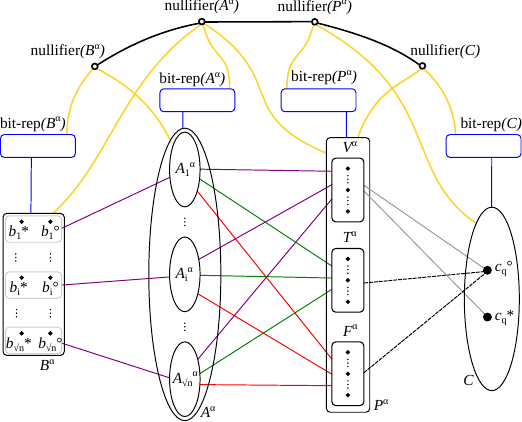}
    \caption{Overview of the reduction.
    Sets in ellipses are independent sets and sets in rectangles are cliques.
    For $X \in \{B^{\alpha}, A^{\alpha},  P^{\alpha}, C\}$,  the blue rectangle attached to it
    via the blue edge represents $\bitrep(X)$.
    We omit $\bits(X)$ for legibility.
    The yellow line represents that $\bitrepnullifier(X)$ is connected to every vertex in the set.
    Note the exception of $\bitrepnullifier(P^{\alpha})$ which is \emph{not} adjacent to any
    vertex in $A^{\alpha}$.
    Purple lines between two sets denote adjacencies with respect to indexing,
    i.e.,  $b^{\alpha, \circ}_i$ is adjacent only with all the vertices in $A^{\alpha}_{i}$,
    and all the vertices in $A^{\alpha}_i$ are adjacent with $v^{\alpha}_i$ in
    validation portal $V^{\alpha}$.
    Gray lines also indicate adjacencies with respect to indexing, but in a complementary
    way.
    If $C_q$ contains a variable in $B^{\alpha}_i$, then $c^{\circ}_q$ and $c^{\star}_q$
    are adjacent with all vertices $v^{\alpha}_j \in V^{\alpha}$ such that $j \neq i$.
    Green and red lines between the $A^{\alpha}$ and $T^{\alpha}$ and $F^{\alpha}$ roughly transfer, for each $a_{i,\ell}^{\alpha}\in A^{\alpha}$, the underlying assignment structure.
    If the $j^{th}$ variable by $a_{i,\ell}^{\alpha}$ is assigned to \true, then we add the green edge $(a_{i,\ell}^{\alpha}, t_{j}^{\alpha})$, and otherwise the red edge $(a_{i,\ell}^{\alpha}, f_{j}^{\alpha})$.
    Similarly, we add edges for each $c_i^{\circ}\in C$ depending on the assignment satisfying the variable from the part $X^{\delta}$ of a clause $c_i$, and in which block $B_j^{\delta}$ it lies, putting either an edge $(c_i^{\circ},t_j^{\delta})$ or $(c_i^{\circ},f_j^{\delta})$ accordingly ($\delta \in \{\alpha, \beta, \gamma\}$).
    }
    \label{fig:reduction-overview}
\end{figure}

\begin{itemize}
\item We partition the variables of $X^{\alpha}$ into \emph{buckets}
$X^{\alpha}_1,  X^{\alpha}_2,  \dots ,X^{\alpha}_{\sqrt{n}}$
such that each bucket contains $\sqrt{n}$ variables.
Let $X^{\alpha}_i = \{x^{\alpha}_{i, j}\ |\ j \in [\sqrt{n}]\}$ for all $i\in [\sqrt{n}]$.

\item For every $X^{\alpha}_i$, we construct a set $A^{\alpha}_i$ of $2^{\sqrt{n}}$ new vertices, $A^{\alpha}_i=\{a^{\alpha}_{i,  \ell}\mid \ell \in [2^{\sqrt{n}}]\}$.
Each vertex in $A^{\alpha}_i$ corresponds to a certain possible assignment of
variables in $X^{\alpha}_i$.
Let $A^{\alpha}$ be the collection of all the vertices added in the above step, that is,
$A^{\alpha} = \{a^{\alpha}_{i, \ell} \in A_i |\ i \in [\sqrt{n}] \text{ and }
\ell \in [2^{\sqrt{n}}]  \}$.
We add a set identifying gadget as mentioned in
Subsection~\ref{subsubsec:gadget-set-MD} in order to resolve every pair of vertices in $A^{\alpha}$.

\item For every $X^{\alpha}_i$, we construct a pair
$\langle b^{\alpha, \circ}_i, b^{\alpha, \star}_i \rangle$ of vertices.
Then, we add a gadget to make the pairs
$\{\langle b^{\alpha, \circ}_i, b^{\alpha, \star}_i \rangle\mid i\in [\sqrt{n}]\}$ critical as mentioned in Subsection~\ref{subsubsec:gadget-critical-pairs}.
Let $B^{\alpha} = \{b^{\alpha, \circ}_i, b^{\alpha, \star}_i\  |\ i \in [\sqrt{n}] \}$
be the collection of vertices in the critical pairs.
We add edges in $B^{\alpha}$ to make it a clique.

\item We would like that
any resolving set contains at least one vertex in $A^{\alpha}_i$
for every $i \in [\sqrt{n}]$, but instead we add the construction from Subsection~\ref{subsubsec:vertex-selector} that achieves the slightly weaker objective as mentioned there.
However,  for every $A^{\alpha}_i$,  instead of adding two new vertices, we use
$\langle b^{\alpha, \circ}_i,  b^{\alpha,  \star}_i \rangle$ as the necessary critical pair.
Formally,  for every $i \in [\sqrt{n}]$,  we make $b^{\alpha, \circ}_i$ adjacent to every vertex
in $A^{\alpha}_{i}$.
We add edges to make $\bitrepnullifier(B^{\alpha})$ adjacent to every vertex
in $A^{\alpha}$, and
$\bitrepnullifier(A^{\alpha})$ adjacent to every vertex in $B^{\alpha}$.
Recall that there is also the edge $(\bitrepnullifier(B^{\alpha}), \bitrepnullifier(A^{\alpha}))$.

\item We add \emph{portals} that transmit information
from vertices corresponding to assignments, i.e., vertices in $A^{\alpha}$,
to critical pairs corresponding to clauses, i.e., vertices in $C$
{which we define soon}.
A portal is a clique on $\sqrt{n}$ vertices in the graph $G$.
We add three portals, the \emph{truth portal} (denoted by $T^{\alpha}$),
\emph{false portal} (denoted by $F^{\alpha}$), and
\emph{validation portal} (denoted by $V^{\alpha}$).
Let
$T^{\alpha} = \{t^{\alpha}_1,  t^{\alpha}_2, \dots, t^{\alpha}_{\sqrt{n}}\}$,
$F^{\alpha} = \{f^{\alpha}_1,  f^{\alpha}_2, \dots, f^{\alpha}_{\sqrt{n}}\}$,
and
$V^{\alpha} = \{v^{\alpha}_1,  v^{\alpha}_2, \dots, v^{\alpha}_{\sqrt{n}}\}$.
Moreover,  let $P^{\alpha} = V^{\alpha} \cup T^{\alpha} \cup F^{\alpha}$.

\item We add a set identifying gadget for $P^{\alpha}$ as mentioned
in Subsection~\ref{subsubsec:gadget-set-MD}.
We add an edge between $\bitrepnullifier(A^{\alpha})$ and every vertex of $P^{\alpha}$; and the edge $(\bitrepnullifier(P^{\alpha}),  \bitrepnullifier(A^{\alpha}))$.
However, we note that we \emph{do not} add edges between
$\bitrepnullifier(P^{\alpha})$ and $A^{\alpha}$, as can be seen in Figure~\ref{fig:reduction-overview}.
Lastly, we add edges in $P^{\alpha}$ to make it a clique.

\item We add edges between $A^{\alpha}$ and the portals as follows.
For $i \in [\sqrt{n}]$ and $\ell \in [2^{\sqrt{n}}]$,
consider a vertex $a^{\alpha}_{i, \ell}$ in $A^{\alpha}_i$.
Recall that this vertex corresponds to an assignment
$\pi: X^{\alpha}_i \mapsto \{\true, \false\}$,
where $X_i^{\alpha}$ is the collection of variables
$\{x^{\alpha}_{i, j}\ |\ j \in [\sqrt{n}] \}$.
If $\pi(x^{\alpha}_{i, j}) = \true$, then we add the edge $(a^{\alpha}_{i,  \ell},  t^{\alpha}_j)$.
Otherwise,  $\pi(x^{\alpha}_{i, j}) = \false$, and we add the edge
$(a^{\alpha}_{i,  \ell},  f^{\alpha}_j)$.
We add the edge $(a^{\alpha}_{i,  \ell},  v^{\alpha}_{i})$ for every $\ell \in [2^{\sqrt{n}}]$.
\end{itemize}

Then, we repeat the above steps to construct
$B^{\beta},  A^{\beta},  P^{\beta}$,
$B^{\gamma},  A^{\gamma},  P^{\gamma}$.

Now, we are ready to proceed through the final steps to complete the construction.
\begin{itemize}
\item For every clause $C_q$ in $\psi$,  we introduce a pair of vertices
$\langle c^{\circ}_q,  c^{\star}_q \rangle$.
Let $C$ be the collection of vertices in such pairs.
Then, we add a gadget as was described in Subsection~\ref{subsubsec:gadget-critical-pairs} to make each pair
$\langle c^{\circ}_q,  c^{\star}_q \rangle$ a critical one.

\item For each $\delta \in \{\alpha, \beta, \gamma\}$, we add an edge between $\bitrepnullifier(P^{\delta})$ and every vertex of $C$, and we add the edge $(\bitrepnullifier(P^{\delta}),  \bitrepnullifier(C))$. Now, we add edges between $C$ and the portals as follows for each $\delta \in \{\alpha, \beta, \gamma\}$.
Consider a clause $C_q$ in $\psi$ and the corresponding
critical pair $\langle c^{\circ}_q,  c^{\star}_q \rangle$ in $C$.
As $\psi$ is an instance of \textsc{$3$-Partitioned-$3$-SAT},
there is at most one variable in $X^{\delta}$ that appears in $C_q$.
If $C_q$ does not contain a variable in $X^{\delta}$,
then we make $c^{\circ}_q$ and $c^{\star}_q$
adjacent to every vertex in $V^{\delta}$, and they
are not adjacent to any vertex in $T^{\delta} \cup F^{\delta}$.
Otherwise, suppose that $C_q$ contains the variable $x^{\delta}_{i, j}$ for some $i, j \in [\sqrt{N}]$.
The first subscript decides the edges between
$\langle c^{\circ}_q,  c^{\star}_q \rangle$ and the validation portal,
whereas the second subscript decides the edges between
$\langle c^{\circ}_q,  c^{\star}_q \rangle$ and
either the truth portal or false portal in the following sense.
We add all edges of the form
$(v^{\delta}_{i'},  c^{\circ}_q)$ and $(v^{\delta}_{i'},  c^{\star}_q)$
for every $i' \in [\sqrt{n}]$ such that $i' \neq i$.
If $x^{\delta}_{i, j}$ appears as a positive literal in $C_q$, then we add the edge $(t^{\delta}_j,  c^{\circ}_q)$.
Otherwise, $x^{\delta}_{i, j}$ appears as a negative literal in $C_q$, and we add the edge $(f^{\delta}_j,  c^{\circ}_q)$.
\end{itemize}

This concludes the construction of $G$.
The reduction returns $(G,  k)$
as an instance of \textsc{Metric Dimension} where
\begin{equation*}
\begin{split}
k &= 3 \cdot \left(\sqrt{n} + \left(\left\lceil\log(|B^{\alpha}|/2 + 2)\right\rceil + 1\right)
+ \left(\left\lceil\log(|A^{\alpha}|+ 2)\right\rceil + 1\right) + \left(\left\lceil\log(|P^{\alpha}|+ 2)\right\rceil + 1\right) \right) +
\\
&  \lceil\log(|C|/2+ 2)\rceil + 1.
\end{split}
\end{equation*}

%\subsection{Correctness of the Reduction}\label{subsec:correctness-3-Par-3-SAT-Met-Dim}

\begin{figure}[t]
    \centering
        \includegraphics[scale=1.25]{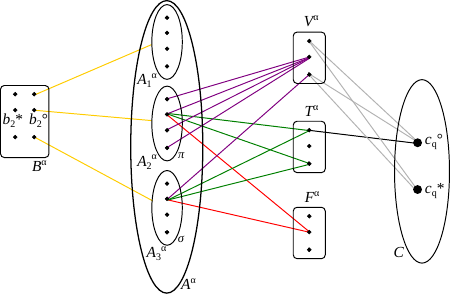}
    \caption{A toy example to illustrate the core ideas in the reduction. Note that $\bitrep$ and $\bitrepnullifier$ are omitted for the sets.}
    \label{fig:reduction-core}
\end{figure}

We give an informal description of the proof of correctness here.
See Figure~\ref{fig:reduction-core}.
Suppose $\sqrt{n} = 3$ and the vertices in the sets are indexed from top to bottom.
For the sake of clarity,  we do not show all the edges and only show $4$ out of $8$ vertices
in each $A^{\alpha}_i$ for $i \in [3]$.
We also omit $\bitrep$ and $\bitrepnullifier$ for these sets.
The vertex selection gadget and the budget $k$ ensure that exactly one vertex in
$\{b^{\alpha,\circ}_i,  b^{\alpha,\star}_i\} \cup A^{\alpha}_i$ is selected for every $i \in [3]$.
If a resolving set contains a vertex from $A^{\alpha}_i$,
then it corresponds to selecting an assignment of variables in $X^{\alpha}_i$.
For example,  the vertex $a^{\alpha}_{2, 2}$ corresponds to the assignment
$\pi: X^{\alpha}_2 \mapsto \{\true, \false\}$.
Suppose $X^{\alpha}_2 = \{x^{\alpha}_{2, 1},  x^{\alpha}_{2, 2},  x^{\alpha}_{2, 3}\}$,
$\pi(x^{\alpha}_{2, 1}) = \pi(x^{\alpha}_{2,3}) = \true$, and $\pi(x^{\alpha}_{2, 2}) = \false$.
Hence,  $a^{\alpha}_{2, 2}$ is adjacent to the first and third vertex in the truth portal $T^{\alpha}$,
whereas it is adjacent with the second vertex in the false portal $F^{\alpha}$.
Suppose the clause $C_q$ contains the variable $x^{\alpha}_{2, 1}$ as a positive literal.
Note that $c^{\circ}_q$ and $c^{\star}_q$ are at distance $2$ and $3$,  respectively,  from
$a^{\alpha}_{2, 2}$.
Hence, the vertex $a^{\alpha}_{2, 2}$, corresponding to the assignment $\pi$ that satisfies clause $C_q$,
resolves the critical pair $\langle c^{\circ}_q,  c^{\star}_q \rangle$.
Now, suppose there is another assignment $\sigma: X^{\alpha}_3 \mapsto \{\true, \false\}$ such that
$\sigma(x^{\alpha}_{3, 1}) = \sigma(x^{\alpha}_{3,3}) = \true$ and $\sigma(x^{\alpha}_{3, 2}) = \false$.
As $\psi$ is an instance of \textsc{$3$-Partitioned-$3$-SAT}
and $C_q$ contains a variable in $X^{\alpha}_2$ ($ \subseteq X^{\alpha}$),
$C_q$ does not contain a variable in $X^{\alpha}_3$ ($\subseteq X^{\alpha}$).
Hence,  $\sigma$ does not satisfy $C_q$.
Let $a^{\alpha}_{3, 2}$ be the vertex in $X^{\alpha}_3$ corresponding to $\sigma$.
The connections via the validation portal $V^{\alpha}$ ensure that
both $c^{\circ}_q$ and $c^{\star}_q$ are at distance $2$ from
$a^{\alpha}_{3, 2}$, and hence, $a^{\alpha}_{3, 2}$ cannot resolve the
critical pair $\langle c^{\circ}_q,  c^{\star}_q \rangle$.
Hence,  finding a resolving set in $G$ corresponds to finding a satisfying
assignment for $\psi$. These intuitions are formalized in the following subsection.

\subsection{Correctness of the Reduction}

Suppose, given an instance $\psi$ of \textsc{$3$-Partitioned-$3$-SAT}, that the reduction of this subsection returns $(G, k)$ as an instance of \textsc{Metric Dimension}. We first prove the following lemma which will be helpful in proving the correctness of the reduction.

\begin{lemma} For any resolving set $S$ of $G$ and for all $X\in \{C\} \cup \{B^{\delta},A^{\delta},P^{\delta} \mid \delta\in \{\alpha,\beta,\gamma\}\}$,
\begin{enumerate}
\item $S$ contains at least one vertex from each pair of false twins in $\bits(X)$.
\item Vertices in $\bits(X) \cap S$ resolve any non-critical pair of vertices
$\langle u, v\rangle$ when $u \in X\cup X^+$ and $v\in V(G)$.  
\item Vertices in $X^+ \cap S$ \emph{cannot} resolve any critical
pair of vertices $\langle b_i^{\delta',\circ},  b_i^{\delta',\star}\rangle$ nor $\langle c_q^{\circ},  c_q^{\star}\rangle$ for all $i\in [\sqrt{n}]$, $\delta'\in \{\alpha,\beta,\gamma\}$, and $q\in [m]$.     
\end{enumerate}\label{claim:set-ind}
\end{lemma}
\begin{proof}
\begin{enumerate}
\item By~\cref{obs:twins}, the statement follows for all $X\in \{C\} \cup \{B^{\delta},A^{\delta},P^{\delta} \mid \delta\in \{\alpha,\beta,\gamma\}\}$.

\item For all $X\in \{C\} \cup \{B^{\delta},A^{\delta},P^{\delta} \mid \delta\in \{\alpha,\beta,\gamma\}\}$, note that $\bitrepnullifier(X)$ is distinguished by $\bits(X)\cap S$ since it is the only vertex in $G$ that is at distance~$2$ from every vertex in $\bits(X)$. We now do a case analysis for the remaining non-critical pairs of vertices $\langle u, v \rangle$ assuming that $\bitrepnullifier(X)\notin \{u,v\}$ (also, suppose that both $u$ and $v$ are not in $S$, as otherwise, they are obviously distinguished):
\begin{description}
\item[Case i: $u, v \in X\cup X^+$.]
\hfill
\begin{description}
\item[Case i(a): $u, v\in X$ or $u, v\in \bitrep(X)\setminus \bits(X)$.] In the first case, let $j$ be the bit in the binary representation of the subscript of $u$ that is not equal to the $j^{\text{th}}$ bit in the binary representation of the subscript of $v$ (such a $j$ exists since $\langle u, v \rangle$ is not a critical pair). In the second case, without loss of generality, let $u=y_i$ and $v=y_j$. 
By the first item of the statement of the lemma (1.), without loss of generality, $y_j^{a}\in S\cap \bits(X)$.
Then, in both cases, $d(y_j^a, u)\neq d(y_j^a, v)$.
\item[Case i(b): $u \in X$ and $v\in \bitrep(X)$.]
Without loss of generality, $y_{\star}^{a}\in S\cap \bits(X)$ (by 1.). 
Then, $d(y_{\star}^{a}, u)=2$ and, for all $v\in \bits(X)\setminus\{y_{\star}^b\}$, $d(y_{\star}^{a}, v)=3$. 
Without loss of generality, let $y_i$ be adjacent to $u$ and let $y_i^a\in S\cap \bits(X)$ (by 1.).
Then, for $v=y_{\star}^b$, $3=d(y_i^a, v)\neq d(y_i^a, u)=2$.
If $v\in \bitrep(X)\setminus \bits(X)$, then, without loss of generality, $v=y_j$ and $y_j^a\in S\cap \bits(X)$ (by 1.), and $1=d(y_j^a, v)< d(y_j^a, u)$.
\item[Case i(c): $u, v\in \bits(X)$.]
Without loss of generality, $u=y_i^b$ and $y_i^a\in S$ (by 1.). 
Then, $2=d(y_i^a, u)\neq d(y_i^a, v)=3$.
\item[Case i(d): $u \in \bits(X)$ and $v\in \bitrep(X)\setminus \bits(X)$.]
Without loss of generality, $v=y_i$ and $y_i^a\in S$ (by 1.). 
Then, $1=d(y_i^a, v) < d(y_i^a, u)$.
\end{description}
\item[Case ii: $u \in X\cup X^+$ and $v \in V(G)\setminus (X\cup X^+)$.]
For each $u \in X\cup X^+$, there exists $w\in \bits(X)\cap S$ such that $d(u, w)\leq 2$, while, for each $v \in V(G)\setminus (X\cup X^+)$ and $w\in \bits(X)\cap S$, we have $d(v, w)\geq 3$.
\end{description}
\item For all $X\in \{B^{\delta},A^{\delta},P^{\delta} \mid \delta\in \{\alpha,\beta,\gamma\}\}$, $u\in X^+$, $v \in \{c_q^{\circ}, c_q^{\star}\}$, and $q\in [m]$,
we have that $d(u, v)=d(u, \bitrepnullifier(P^{\delta}))+1$.
Further, for $X=C$ and all $u\in X^+$ and $q\in [m]$, either $d(u,c_q^{\circ})=d(u,c_q^{\star})=1$, $d(u,c_q^{\circ})=d(u,c_q^{\star})=2$, or $d(u,c_q^{\circ})=d(u,c_q^{\star})=3$ by the construction in Subsection~\ref{subsubsec:gadget-critical-pairs} and since $\bitrep(X)\setminus \bits(X)$ is a clique. 
Hence, for all $X\in \{C\} \cup \{B^{\delta},A^{\delta},P^{\delta} \mid \delta\in \{\alpha,\beta,\gamma\}\}$, vertices in $X^+ \cap S$ cannot resolve a pair of vertices $\langle c_q^{\circ}, c_q^{\star}\rangle$ for any $q\in [m]$.

\smallskip

For all $\delta\in \{\alpha,\beta,\gamma\}$, if $v\in B^{\delta}$, then, for all $X\in \{C\} \cup \{B^{\delta'},A^{\delta'},P^{\delta'} \mid \delta'\in \{\alpha,\beta,\gamma\}\}$ such that $\delta\neq \delta'$, and $u\in X^+$, we have that $d(u,v)=d(u, \bitrepnullifier(A^{\delta}))+1$. Similarly, for all $\delta\in \{\alpha,\beta,\gamma\}$, if $v\in B^{\delta}$, then, for all $X\in \{A^{\delta}, P^{\delta}\}$ and $u\in X^+$, we have that $d(u,v)=d(u, \bitrepnullifier(A^{\delta}))+1$. Lastly, for each $\langle b_i^{\delta,\circ},  b_i^{\delta,\star}\rangle$, $\delta\in \{\alpha,\beta,\gamma\}$, and $i\in [\sqrt{n}]$, if $X=B^{\delta}$, then, for all $u\in X^+$, either $d(u,b_i^{\delta,\circ})=d(u,b_i^{\delta,\star})=1$, $d(u,b_i^{\delta,\circ})=d(u,b_i^{\delta,\star})=2$, or $d(u,b_i^{\delta,\circ})=d(u,b_i^{\delta,\star})=3$ by the construction in Subsection~\ref{subsubsec:gadget-critical-pairs} and since $\bitrep(X)\setminus \bits(X)$ is a clique. \qedhere
\end{enumerate}
\end{proof}

\begin{lemma}
\label{lemma:3-Part-3-SAT-Met-Dim-forward}
If $\psi$ is a satisfiable \textsc{$3$-Partitioned-$3$-SAT} formula,  then
$G$ admits a resolving set of size $k$.
\end{lemma}
\begin{proof}
Suppose $\pi: X^{\alpha} \cup X^{\beta} \cup X^{\gamma} \mapsto \{\true, \false\}$
is a satisfying assignment for $\psi$.
We construct a resolving set $S$ of size $k$ for $G$ using this assignment.

Initially, set $S = \emptyset$.
For every $\delta \in \{\alpha, \beta, \gamma\}$ and $i \in [\sqrt{n}]$,
consider the assignment $\pi$ restricted to the variables in $X^{\delta}_i$.
By the construction,  there is a vertex in $A^{\delta}_i$ that corresponds
to this assignment.
Include that vertex in $S$.
For each $X \in \{C\} \cup \{B^{\delta},  A^{\delta},  P^{\delta} \mid \delta \in \{\alpha, \beta, \gamma\}\}$,
we add one vertex from each pair of the false twins in $\bits(X)$ to $S$.
Note that $|S| = k$ and that every vertex in $S$ is distinguished by itself.

In the remaining part of the proof, we show that $S$ is a resolving set of $G$.
First, we prove that all critical pairs are resolved by $S$ in the following claim.

\begin{claim}\label{clm:critpairs}
All critical pairs are resolved by $S$.
\end{claim}
\begin{claimproof}
For each $i\in [\sqrt{n}]$ and $\delta\in \{\alpha, \beta, \gamma\}$, the critical pair $\langle b^{\delta, \circ}_i, b^{\delta, \star}_i \rangle$
is resolved by the vertex $S\cap A_i^{\delta}$ by the construction.
For each $q\in [m]$, the clause $C_q$ is satisfied by the assignment $\pi$.
Thus, there is a variable $x$ in $C_q$ that satisfies $C_q$ according to $\pi$.
Suppose that $x\in X_i^{\delta}$.
Let $a_{i,\ell}^{\delta}$ be the vertex in $A_i^{\delta}$ corresponding to $\pi$.
Then, by the construction, $d(a_{i,\ell}^{\delta},c_q^{\circ})=2<3=d(a_{i,\ell}^{\delta},c_q^{\star})$.
Thus, every critical pair $\langle c_q^{\circ}, c_q^{\star} \rangle$ is resolved by $S$.
\end{claimproof}

Then, every vertex pair in $V(G)$ is resolved by $S$ by Claim~\ref{clm:critpairs} in conjunction with the second item of the statement of \cref{claim:set-ind}.
\end{proof}

\begin{lemma}
\label{lemma:3-Part-3-SAT-Met-Dim-backword}
If $G$ admits a resolving set of size $k$, then $\psi$ is a satisfiable \textsc{$3$-Partitioned-$3$-SAT} formula.
\end{lemma}
\begin{proof}
Assume that $G$ admits a resolving set $S$ of size $k$.
First, we prove some properties of~$S$.
%Recall that, as mentioned in Subsection~\ref{subsubsec:gadget-set},
%there is a pair of false twins for each $y\in\bitrep(X)\setminus \bits(X)$
%for each $X \in \{B^{\delta},  A^{\delta},  P^{\delta}, C\}$ and
%$\delta \in \{\alpha, \beta, \gamma\}$.
%Based on Observation~\ref{obs:twins}, for any resolving set $S$, at least one vertex from each pair of twins is in $S$.
By the first item of the statement of \cref{claim:set-ind}, for each $\delta\in \{\alpha, \beta, \gamma\}$, we have that
\begin{align*}
|S\cap \bits(A^{\delta})| \geq \lceil\log(|A^{\delta}|+2)\rceil + 1,\qquad\, 
 &|S\cap \bits(P^{\delta})| \geq \lceil\log(|P^{\delta}|+2)\rceil + 1, \\
|S\cap \bits(B^{\delta})| \geq \lceil\log(|B^{\delta}|/2+2)\rceil + 1, \quad
&|S\cap \bits(C)| \geq \lceil\log(|C|/2+2)\rceil + 1.
\end{align*}
Hence, any resolving set $S$ of $G$ already has size at least $$3 \cdot \left(\left(\left\lceil\log(|B^{\alpha}|/2 + 2)\right\rceil + 1\right)
+ \left(\left\lceil\log(|A^{\alpha}|+ 2)\right\rceil + 1\right) + \left(\left\lceil\log(|P^{\alpha}|+ 2)\right\rceil + 1\right) \right) + \lceil\log(|C|/2+ 2)\rceil + 1.$$

Now, for each $\delta\in\{\alpha, \beta, \gamma\}$ and $i\in[\sqrt{n}]$, consider the critical pair $\langle b_i^{\delta, \circ}, b_i^{\delta, \star} \rangle$.
By the construction mentioned in Subsection~\ref{subsubsec:gadget-critical-pairs}, only $v\in A_i^{\delta}\cup \{b_i^{\delta, \circ}, b_i^{\delta, \star}\}$ resolves a pair $\langle b_i^{\delta, \circ}, b_i^{\delta, \star} \rangle$.
Indeed, for all $X\in \{C\} \cup \{B^{\delta'},A^{\delta'},P^{\delta'} \mid \delta'\in \{\alpha,\beta,\gamma\}\}$, no vertex in $X^+$ can resolve such a pair by the third item of the statement of \cref{claim:set-ind}. Also, for all $X\in \{C\}\cup \{A^{\delta''},P^{\delta'} \mid \delta' \in \{\alpha, \beta, \gamma\}, \delta'' \in \{\alpha, \beta, \gamma\}\}$ such that $\delta\neq \delta''$, and $u\in X$, we have that $d(u,b_i^{\delta, \circ})=d(u,b_i^{\delta, \star})=d(u,\bitrepnullifier(A^{\delta}))+1$. Furthermore, for any $a\in A_j^{\delta}$ with $j\in [\sqrt{n}]$ such that $i\neq j$, we have that $d(a,b_i^{\delta, \circ})=d(a,b_i^{\delta, \star})=2$ by construction and since $B^{\delta}$ is a clique. 
Hence, since any resolving set $S$ of $G$ of size at most $k$ can only admit at most another $3\sqrt{n}$ vertices, we get that equality must in fact hold in every one of the aforementioned inequalities, and any resolving set $S$ of $G$ of size at most $k$ contains one vertex from $A_i^{\delta}\cup \{b_i^{\delta, \circ}, b_i^{\delta, \star}\}$ for all $i\in [\sqrt{n}]$ and $\delta\in \{\alpha, \beta, \gamma\}$. Hence, any resolving set $S$ of $G$ of size at most $k$ is actually of size exactly $k$.

Next, for each $\delta\in \{\alpha, \beta, \gamma\}$, we construct an assignment $\pi: X^{\alpha} \cup X^{\beta} \cup X^{\gamma} \mapsto \{\true, \false\}$ in the following way.
If $a_{i,\ell}^{\delta}\in S$ and $\pi':X_{i}^{\delta}\rightarrow \{\true, \false\}$ corresponds to the underlying assignment of $a_{i,\ell}^{\delta}$ for the variables in $X_{i}^{\delta}$,
then let $\pi:X_{i}^{\delta}\rightarrow \{\true, \false\}:=\pi'$ for each $i\in [\sqrt{n}]$ and $\delta\in \{\alpha, \beta, \gamma\}$.
If $S\cap A_i^{\delta}=\emptyset$, then one of $ b_i^{\delta, \circ}, b_i^{\delta, \star}$ is in $S$, and we can use an arbitrary assignment of the variables in the bucket $X_i^{\delta}$.

We now prove that the constructed assignment $\pi$ satisfies every clause in $C$.
Since $S$ is a resolving set, it follows that, for every clause $c_q\in C$,
there exists $v\in S$ such that $d(v, c_q^{\circ})\neq d(v, c_q^{\star})$.
Notice that, for any $v$ in $\bits(A^{\delta}), \bits(B^{\delta}), \bits(P^{\delta})$ for any $\delta\in \{\alpha, \beta, \gamma\}$ or in $\bits(C)$, we have $d(v, c_q^{\circ})= d(v, c_q^{\star})$ by the third item of the statement of Lemma~\ref{claim:set-ind}.
Further, for any $v\in B^{\delta}$ and any $\delta\in \{\alpha, \beta, \gamma\}$, we have that $d(v, c_q^{\circ})=d(v, c_q^{\star})=d(v,\bitrepnullifier(P^{\delta}))+1$.
Thus, $v\in S\cap \bigcup\limits_{\delta\in \{\alpha, \beta, \gamma\}} A^{\delta}$.
Without loss of generality, suppose that $c_q^{\circ}$ and $c_q^{\star}$ are resolved by
$a_{i,\ell}^{\alpha}$.
So, $d(a_{i,\ell}^{\alpha}, c_q^{\circ}) \neq d(a_{i,\ell}^{\alpha}, c_q^{\star})$.
By the construction, the only case where $d(a_{i,\ell}^{\alpha}, c_q^{\circ}) \neq d(a_{i,\ell}^{\alpha}, c_q^{\star})$ is when
$C_q$ contains a variable $x\in X_i^{\alpha}$ and $\pi(x)$ satisfies $C_q$.
Thus, we get that the clause $C_q$ is satisfied by the assignment $\pi$.
%Then, the shortest path from $a_{i,l}^{\alpha}$ to $c_q^{\circ}$ passes through the only vertex $w\in P_T^{\alpha}\cup P_F^{\alpha}$ that is in the neighborhood of $c_i^{\circ}$ and not in the neighborhood of $c_q^{\circ}$.
%Thus $d(a_{i,l}^{\alpha}, c_q^{\circ})=2$.
%See, that the neighborhoods of $c_i^{\circ}$ and $c_i^{\star}$ differ by only one vertex, since we have a valid \textsc{3P-3SAT} instance.
%Since $d(a_{i,l}^{\alpha}, c_q^{\star})>2$ by the construction,
%the only way to get shorter path from $a_{i,l}^{\alpha}$ to $c_i^{\circ}$ is through $w\in P_T^{\alpha}\cup P_F^{\alpha}$ with the distance $2$.
%But this path exists in the construction, if the underlying assignment of $a_{i,l}^{\alpha}$ for variables $B_{i}^{\alpha}$ assign the same value which satisfies the clause $c_j$.

Since $S$ resolves all pairs $\langle c_q^{\circ},c_q^{\star}\rangle$ in $V(G)$, the assignment $\pi$ constructed above indeed satisfies every clause $c_q$, completing the proof.
\end{proof}

\begin{proof}[Proof of Theorem~\ref{thm:reduction-3-SAT-Met-Dim-VC}.]
The proof of \cref{prop:3-SAT-to-3-Partition-3-SAT-lampis}, 
relies on the fact that there is a polynomial-time reduction from 
\textsc{3-SAT} to \textsc{$3$-Partitioned-$3$-SAT} that increases the number of variables and clauses by a constant factor.
In Subsection~\ref{subsec:reduction-3-Par-3-SAT-Met-Dim}, we presented a reduction that takes an instance $\psi$ of \textsc{$3$-Partitioned-$3$-SAT} and returns an equivalent instance $(G,k)$ of \textsc{Metric Dimension} (by Lemmas~\ref{lemma:3-Part-3-SAT-Met-Dim-forward} and \ref{lemma:3-Part-3-SAT-Met-Dim-backword}) in $2^{\OO(\sqrt{n})}$ time, where
\begin{equation*}
\begin{split}
k &= 3 \cdot \left(\sqrt{n} + \left(\left\lceil\log(|B^{\alpha}|/2 + 2)\right\rceil + 1\right)
+ \left(\left\lceil\log(|A^{\alpha}|+ 2)\right\rceil + 1\right) + \left(\left\lceil\log(|P^{\alpha}|+ 2)\right\rceil + 1\right) \right) +
\\
&  (\lceil\log(|C|/2+ 2)\rceil + 1)=\OO(\sqrt{n}).
\end{split}
\end{equation*} 
Note that $V(G)=2^{\OO(\sqrt{n})}$. 
%Then, note that the diameter of the graph $G$ is $4$.
Further, note that taking all the vertices in $B^{\delta}$ and $P^{\delta}$ for all $\delta\in \{\alpha,\beta,\gamma\}$, and $X^+\setminus \bits(X)$ for all $X\in \{C\} \cup \{B^{\delta},A^{\delta},P^{\delta} \mid \delta\in \{\alpha,\beta,\gamma\}\}$, results in a vertex cover of $G$. Hence, 
\begin{equation*}
\begin{split}
\vc(G) &\leq 3 \cdot \left(\left(\left\lceil\log(|B^{\alpha}|/2 + 2)\right\rceil + 2\right)
+ \left(\left\lceil\log(|A^{\alpha}|+ 2)\right\rceil + 2\right) + \left(\left\lceil\log(|P^{\alpha}|+ 2)\right\rceil + 2\right)\right) +
\\
& 3\cdot (|B^{\alpha}|+|P^{\alpha}|) +  (\lceil\log(|C|/2+ 2)\rceil + 2)=\OO(\sqrt{n}).
\end{split}
\end{equation*}

Thus, $\vc(G)+k=\OO(\sqrt{n})$.
\end{proof}

% !TEX root = ./main.tex
\section{\gsfull: Lower Bounds Regarding Vertex Cover}
\label{sec:lower-bound-vertex-cover-GS}

In this section, we follow the same
template as in Section~\ref{sec:lower-bound-vertex-cover} and first prove
the following theorem.

\begin{theorem}
\label{thm:reduction-3-SAT-GS-VC}
There is an algorithm that, given an instance
$\psi$ of \textsc{3-Partitioned-3-SAT} on $N$ variables,  
runs in $2^{\calO(\sqrt{N})}$ time, and constructs an equivalent 
instance $(G, k)$ of \gsfull such that 
$\vc(G) + k = \OO(\sqrt{N})$ (and $|V(G)| = 2^{\OO(\sqrt{N})}$).
\end{theorem}

The proofs of the following two
corollaries are analogous to the ones for \mdfull.

\begin{corollary}
\label{cor:GS-vc-lower-bound-algo}
Unless the \ETH\ fails, 
\gsfull does not admit an algorithm 
running in $2^{o(\vc^2)} \cdot n^{\OO(1)}$ time. 
\end{corollary}
\begin{corollary}
\label{cor:GS-vc-lower-bound-kernel}
Unless the \ETH\ fails, 
\gsfull\ does not admit a kernelization algorithm that
does not increase the solution size $k$ and
outputs a kernel with $2^{o(k + \vc)}$ vertices. 
\end{corollary}

\subsection{Reduction}
\label{subsec:reduction-3-Par-3-SAT-Geod-Set}

Consider an instance $\psi$ of \textsc{3-Partitioned-3-SAT}
with $X^{\alpha}, X^{\beta}, X^{\gamma}$ the partition of the variable set, where $|X^{\alpha}| = |X^{\beta}| = |X^{\gamma}|=n$. 
By adding dummy variables in each of these sets, we can assume that $\sqrt{n}$ is an integer.
Further, let $\calC=\{C_1, \dots, C_m\}$ be the set of all the clauses of $\psi$.
From $\psi$, we construct the graph $G$ as follows.
We describe the construction for the part of the graph $G$ corresponding to $X^{\alpha}$, with the parts corresponding to $X^{\beta}$ and $X^{\gamma}$ being analogous.
We rename the variables in $X^{\alpha}$
to $x^{\alpha}_{i, j}$ for $i, j \in [\sqrt{n}]$.

\begin{figure}[t]
    \centering
        \includegraphics[scale=1.25]{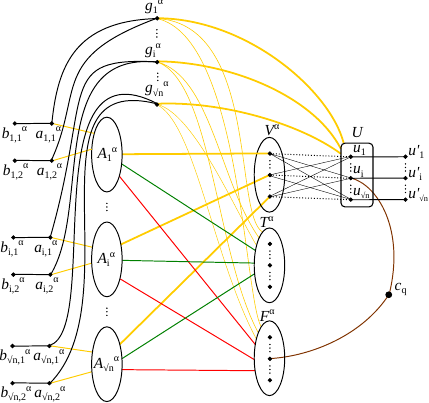}
    \caption{Overview of the reduction.
    Sets with elliptical boundaries are independent sets, and sets with rectangular boundaries are cliques.
    For each $\delta \in \{\alpha, \beta, \gamma\}$, the sets $V^{\delta}$ and $U$ almost form a complete bipartite graph, except for the matching (marked by dotted edges) that is excluded.
    Yellow lines from a vertex to a set denote that this vertex is connected to all the vertices in that set.
    The green and red lines between the $A_i^{\alpha}$ and $T^{\alpha}\cup F^{\alpha}$ transfer, in some sense, for each $w\in A_i^{\alpha}$, the underlying assignment structure.
    If an underlying assignment $w$ sets the $j^{th}$ variable to \true, then we add the green edge $(w,t_{j}^{\alpha})$, and otherwise, we add the red edge $(w,f_{j}^{\alpha})$.
    For all $q \in [m]$ and $\delta \in \{\alpha, \beta, \gamma\}$, let $x_{i, j}^{\delta}$ be the variable in $X^{\delta}$ that is contained in the clause $C_q$ in $\psi$. So, for all $q\in [m]$, if assigning \true\ (\false, respectively) to $x_{i, j}^{\delta}$ satisfies $C_q$, then we add the edge $(c_q,t_j^{\delta})$ ($(c_q,f_j^{\delta})$, respectively).}
    \label{fig:reduction-vc}
\end{figure}

\begin{itemize}
\item We partition the variables of $X^{\alpha}$ into \emph{buckets}
$X^{\alpha}_1,  X^{\alpha}_2,  \dots , X^{\alpha}_{\sqrt{n}}$
such that each bucket contains $\sqrt{n}$ many variables.
Let $X^{\alpha}_i = \{x^{\alpha}_{i, j}\ |\ j \in [\sqrt{n}]\}$ for all $i \in [\sqrt{n}]$.

\item For every bucket $X^{\alpha}_i$, we add an independent set $A^{\alpha}_i$ of $2^{\sqrt{n}}$ new vertices, and we add two isolated edges $(a^{\alpha}_{i, 1},b^{\alpha}_{i, 1})$ and $(a^{\alpha}_{i, 2},b^{\alpha}_{i, 2})$. Let $B^{\alpha}=\{a^{\alpha}_{i, j},b^{\alpha}_{i, j} \mid i\in [\sqrt{n}],j\in \{1,2\}\}$.
For all $i \in [\sqrt{n}]$ and $u \in A_i^{\alpha}$, we make both $a^{\alpha}_{i,1}$ and $a^{\alpha}_{i,2}$ adjacent to $u$ (see \cref{fig:reduction-vc}).

Each vertex in $A^{\alpha}_i$ corresponds to a certain possible assignment of
variables in $X^{\alpha}_i$. 
\item Then, we add three independent sets $T^{\alpha}$, $F^{\alpha}$, and $V^{\alpha}$ on $\sqrt{n}$ vertices each: $T^{\alpha}=\{t^{\alpha}_i \mid i\in [\sqrt{n}]\}$, $F^{\alpha}=\{f^{\alpha}_i \mid i\in [\sqrt{n}]\}$, and $V^{\alpha}=\{v^{\alpha}_i \mid i\in [\sqrt{n}]\}$. 
\item For each $i\in [\sqrt{n}]$, we connect $v^{\alpha}_i$ with all the vertices in $A^{\alpha}_i$.
\item For each $i \in [\sqrt{n}]$, we add edges between $A^{\alpha}_i$ and $T^{\alpha}$ and between $A^{\alpha}_i$ and $F^{\alpha}$ as follows.
Consider a vertex $w\in A^{\alpha}_i$.
Recall that this vertex corresponds to an assignment
$\pi: X^{\alpha}_i \mapsto \{\true, \false\}$,
where $X_i^{\alpha}$ is the collection of variables
$\{x^{\alpha}_{i, j}\ |\ j \in [\sqrt{n}] \}$.
If $\pi(x^{\alpha}_{i, j}) = \true$, then we add the edge $(w,  t^{\alpha}_j)$, and
otherwise, we add the edge $(w,  f^{\alpha}_j)$.
\item For each $i\in [\sqrt{n}]$, we add a special vertex $g^{\alpha}_i$ (also referred to as a $g$-vertex later on) that is adjacent to each vertex in $T^{\alpha}\cup F^{\alpha}$. Further, $g^{\alpha}_i$ is also adjacent to both $a^{\alpha}_{i, 1}$ and $a^{\alpha}_{i, 2}$ (see \cref{fig:reduction-vc}).
\end{itemize}
This finishes the first part of the construction. 
The second step is to connect the three previously constructed parts for $X^{\alpha}$, $X^{\beta}$, and $X^{\gamma}$.
\begin{itemize}
\item We introduce a vertex set $U=\{u_i \mid i\in [\sqrt{n}]\}$ that forms a clique. Then, for each $u_i$, we add an edge to a new vertex $u'_i$.
Thus, we have a matching $\{(u_i,u'_i) \mid i\in [\sqrt{n}]\}$.
Let $U'=\{u'_i\mid i\in [\sqrt{n}]\}$.
\item For each $\delta\in \{\alpha, \beta, \gamma\}$, we add edges so that the vertices of $U\cup V^{\delta}$ almost form a complete bipartite graph, i.e., $E(G)$ contains edges between all pairs $\langle v,w \rangle$ where $v\in U$ and $w\in V^{\delta}$, except for the matching $\{(v^{\delta}_i,u_i) \mid i\in [\sqrt{n}]\}$.
\item For each $\delta\in \{\alpha, \beta, \gamma\}$ and $i\in [\sqrt{n}]$, we make $g_i^{\delta}$ adjacent to each vertex in $U$.
\item For each $C_q\in \calC$, we add a new vertex $c_q$. Let $C=\{c_q \mid q\in [m]\}$. 
Since we are considering an instance of \PSAT, for each $\delta\in \{\alpha, \beta, \gamma\}$, there is at most one variable in $C_q$ that lies in $X^{\delta}$. If there is one, then without loss of generality, let it be $x_{i,j}^{\delta}$ and do the following.
Make $c_q$ adjacent to $u_{i}$ and if $x_{i,j}^{\delta}=\true$ ($x_{i,j}^{\delta}=\false$, respectively) satisfies $C_q$, then $(c_q,t^{\delta}_j)\in E(G)$ ($(c_q,f^{\delta}_j)\in E(G)$, respectively).
\end{itemize}

This concludes the construction of $G$.
The reduction returns $(G,  k)$
as an instance of \textsc{Geodetic Set} where $k = 10\sqrt{n}$.

\subsection{Correctness of the Reduction}

Suppose, given an instance $\psi$ of \textsc{$3$-Partitioned-$3$-SAT}, that the reduction above returns $(G, k)$ as an instance of \textsc{Geodetic Set}. We first prove the following lemmas which will be helpful in proving the correctness of the reduction, and note that we use distances between vertices to prove that certain vertices are not contained in shortest paths.

\begin{lemma}\label{lem:S'crit}
For all $\delta,\delta'\in \{\alpha, \beta, \gamma\}$, the shortest paths between any two vertices in $B^{\delta}\cup U\cup U'$ do not cover any vertices in $C$ nor $V^{\delta'}$.
\end{lemma}

\begin{proof}
Since, for all $i\in [\sqrt{n}]$, $j\in \{1,2\}$, and $\delta\in \{\alpha, \beta, \gamma\}$, the shortest path from $b_{i, j}^{\delta}$ ($u'_i$, respectively) to any other vertex in $G$ first passes through $a_{i, j}^{\delta}$ ($u_i$, respectively), it suffices to prove the statement of the lemma for the shortest paths between any two vertices in $U\cup (B^{\delta}\setminus \{b_{i,j}^{\delta} \mid i\in [\sqrt{n}],j\in \{1,2\}\})$. 

For all $i\in [\sqrt{n}]$, $j\in \{1,2\}$, and $\delta,\delta'\in \{\alpha, \beta, \gamma\}$, we have $d(a_{i, 1}^{\delta},a_{i, 2}^{\delta})=2$ and $d(a_{i, j}^{\delta},w)\geq 2$ for all $w\in V^{\delta'}\cup C$. For all $i,i'\in [\sqrt{n}]$, $j,j'\in \{1,2\}$, and $\delta,\delta',\delta''\in \{\alpha, \beta, \gamma\}$ such that $i\neq i'$ and/or $\delta\neq \delta'$ (i.e., we are not in the previous case), we have $d(a_{i, j}^{\delta},a_{i', j'}^{\delta'})=4$, $d(a_{i, j}^{\delta},w)=d(a_{i', j'}^{\delta'},w)=3$ for all $w\in C$, and, for any $w'\in V^{\delta''}$, we have $d(a_{i, j}^{\delta},w')\geq 2$, $d(a_{i', j'}^{\delta'},w')\geq 2$, and $3\in \{d(a_{i, j}^{\delta},w'),d(a_{i', j'}^{\delta'},w')\}$.
Further, for any $i, i'\in [\sqrt{n}]$ with $i\neq i'$, $d(u_i,u_{i'})=1$.
Lastly, for all $i, i'\in [\sqrt{n}]$, $j\in \{1,2\}$, and $\delta,\delta'\in \{\alpha, \beta, \gamma\}$, we have $d(a_{i,j}^{\delta},u_{i'})=2$, while $d(a_{i,j}^{\delta},w)\geq 2$ for all $w\in V^{\delta'}\cup C$. Hence, the vertices in $C$ and $V^{\delta'}$ are not covered by any shortest path between any of these pairs.
\end{proof}

\begin{lemma}\label{lem:forceA}
For all $i\in [\sqrt{n}]$ and $\delta\in \{\alpha, \beta, \gamma\}$, $v_i^{\delta}$ can only be covered by a shortest path from a vertex in $A_i^{\delta} \cup \{v_i^{\delta}\}$ to another vertex in $G$.
\end{lemma}

\begin{proof}
As stated in the proof of \cref{lem:S'crit}, we do not need to consider any shortest path with an endpoint that is a degree-$1$ vertex.
First, we show that $v_i^{\delta}$ cannot be covered by a shortest path with one endpoint in $U$ and the other not in $A_i^{\delta} \cup \{v_i^{\delta}\}$. To cover this case, by \cref{lem:S'crit}, we just need to consider all shortest paths between a vertex $w\in U$ and any other vertex $z\in V(G)\setminus (U\cup U'\cup B^{\alpha} \cup B^{\beta} \cup B^{\gamma} \cup A_i^{\delta} \cup \{v_i^{\delta}\})$.
Note that $d(z,v_i^{\delta})\geq 2)$, and so, if $d(w,z)\leq 2$, then $v_i^{\delta}$ is not covered by the pair $\langle w,z \rangle$.
Further, $d(w,w')\leq 3$ for all $w'\in V(G)$, and, for all $z'\in V(G)$ such that $d(w,z')=3$, we have that $d(z',v_i^{\delta})\geq 3$.
Hence, $v_i^{\delta}$ cannot be covered by a shortest path with one endpoint in $U$ and the other not in $A_i^{\delta} \cup \{v_i^{\delta}\}$.

For all $i\in [\sqrt{n}]$ and $\delta\in \{\alpha, \beta, \gamma\}$, we have that $N(v_i^{\delta})=A^{\alpha}_i\cup (U\setminus \{u_i\})$, and $d(w,z)\leq 2$ for any $w,z\in N(v_i^{\delta})$. Hence, for a shortest path between two vertices in $V(G)\setminus (A_i^{\delta} \cup \{v_i^{\delta}\})$ to contain $v_i^{\delta}$, that path must also contain two vertices from $N(v_i^{\delta})$. Furthermore, since $v_i^{\delta}$ cannot be covered by a shortest path with one endpoint in $U$ and the other not in $A_i^{\delta} \cup \{v_i^{\delta}\}$, any shortest path whose endpoints are in $V(G)\setminus (A_i^{\delta} \cup \{v_i^{\delta})\}$ that could cover $v_i^{\delta}$ cannot have any of its endpoints in $N(v_i^{\delta})$, and thus, must have length at least $4$. In particular, this proves the following property that we put as a claim to make reference to later.

\begin{claim}\label{clm:neighbors-v}
For any shortest path whose endpoints cannot be in $N[v_i^{\delta}]$, if its first and second endpoints are at distance at least $\ell_1$ and $\ell_2$, respectively, from any vertex in $N(v_i^{\delta})$, then this shortest path cannot cover $v_i^{\delta}$ if it has length less than $\ell_1+\ell_2+2$.
\end{claim} 

We finish with a case analysis of the possible pairs $\langle u,v \rangle$ to prove that no such shortest path covering $v_i^{\delta}$ exists using \cref{clm:neighbors-v}. We note that, by the arguments above, we do not need to consider the case where $u$ and/or $v$ is a degree-$1$ vertex, nor the case where both $u$ and $v$ are in $U\cup B^{\alpha} \cup B^{\beta} \cup B^{\gamma}$, nor the case where one of $u$ and $v$ is in $U$. We now proceed with the case analysis assuming that $u,v\notin N[v_i^{\delta}]$.

\begin{itemize}
\item \textbf{Case 1:} $\boldsymbol{u\in A_{i'}^{\delta'}}$ \textbf{for any} $\mathbf{i'\in [\sqrt{n}]}$ \textbf{and} $\boldsymbol{\delta'\in \{\alpha, \beta, \gamma\}}$ \textbf{such that} $\boldsymbol{A_{i'}^{\delta'}\neq A_{i}^{\delta}}$\textbf{.}
First, $u$ is at distance at least $2$ from any vertex in $N(v_i^{\delta})$, and hence, if $d(u,v)\leq 4$, then we are done by \cref{clm:neighbors-v}. Note that $d(u,v)\leq 4$ as long as $v\notin A_{i'}^{\delta''}$ for any $\delta''\in \{\alpha, \beta, \gamma\}$ such that $\delta''\neq \delta'$, since $u$ is at distance $2$ from every vertex in $U\setminus \{u_{i'}\}$. However, in the case where $v\in A_{i'}^{\delta''}$, we have that $v$ is also at distance at least $2$ from any vertex in $N(v_i^{\delta})$, and so, since $d(u,v)\leq 5$, we are done by \cref{clm:neighbors-v}.

\item \textbf{Case 2:} $\mathbf{u\in C}$ \textbf{or} $\mathbf{u}$ \textbf{is a} $\mathbf{g}$\textbf{-vertex.}
By \cref{clm:neighbors-v} and the previous case, it suffices to note that $d(u,v)\leq 3$ as long as $v\notin A_{i'}^{\delta'}$ for any $i'\in [\sqrt{n}]$ and $\delta'\in \{\alpha, \beta, \gamma\}$.

\item \textbf{Case 3:} $\boldsymbol{u\in T^{\delta'}\cup F^{\delta'}}$ \textbf{for any} $\boldsymbol{\delta'\in \{\alpha, \beta, \gamma\}}$ \textbf{such that} $\mathbf{\boldsymbol{\delta}'\neq \boldsymbol{\delta}}$\textbf{.}
Since $u$ is at distance $2$ from any vertex in $U$, we have that $d(u,v)\leq 4$, and we are done by \cref{clm:neighbors-v}.

\item \textbf{Case 4:} $\boldsymbol{u\in T^{\delta}\cup F^{\delta}}$\textbf{.}
First, if $v\notin A_{i'}^{\delta'}$ for any $i'\in [\sqrt{n}]$ and $\delta'\in \{\alpha, \beta, \gamma\}$, and $v$ is a vertex with a superscript $\delta$, then $d(u,v)\leq 3$. Otherwise, the path from $u$ to $v$ contains a vertex in $U\cup C$, and thus, does not cover $v_i^{\delta}$ since $d(u,v_i^{\delta})\geq 2$, $u$ is at distance~$2$ (at most $3$, respectively) from any vertex in $U$ ($C$, respectively), and $v_i^{\delta}$ is at distance at least~$2$ from any vertex in $C$. Thus, we are done by the previous cases and \cref{clm:neighbors-v}.

\item \textbf{Case 5:} $\boldsymbol{u\in V^{\delta'}}$ \textbf{for any} $\boldsymbol{\delta'\in \{\alpha, \beta, \gamma\}}$\textbf{.}
For any $\delta''\in \{\alpha, \beta, \gamma\}$, if $v\in B^{\delta''}\cup V^{\delta''}$, then $d(u,v)\leq 3$, and we are done by the previous cases and \cref{clm:neighbors-v}. \qedhere
\end{itemize}
\end{proof}

\begin{lemma}\label{lem:GS-vc-forward}
If $G$ admits a geodetic set of size $k$, then $\psi$ is a satisfiable \textsc{$3$-Partitioned-$3$-SAT} formula.
\end{lemma}

\begin{proof}
Assume that $G$ admits a geodetic set $S$ of size $k$.
Then, let us consider the set $S'=\{u_i'\mid i\in [\sqrt{n}]\}\cup \{b^{\delta}_{i,1}, b^{\delta}_{i,2}\mid \delta\in \{\alpha, \beta, \gamma\}, i\in [\sqrt{n}]\}$ of all the degree-1 vertices in $G$.
By \cref{obs:simplicial}, $S'\subseteq S$. By \cref{lem:forceA}, for each $i\in [\sqrt{n}]$ and $\delta\in \{\alpha, \beta, \gamma\}$, there is at least one vertex from $A_i^{\delta}\cup \{v_i^{\delta}\}$ in $S$.
Since $|S'|=7\sqrt{n}$ and $k=10\sqrt{n}$, for each $i\in [\sqrt{n}]$ and $\delta\in \{\alpha, \beta, \gamma\}$, $S$ contains exactly $1$ vertex from $A_i^{\delta}\cup \{v_i^{\delta}\}$.

By \cref{lem:S'crit}, the shortest paths between vertices in $S'$ do not cover vertices in $C$, and thus, the $3\sqrt{n}$ vertices in $S\setminus S'$ must cover them.

For this goal, the vertices in $S$ that are in $V^{\delta}$ for any $\delta\in \{\alpha, \beta, \gamma\}$ are irrelevant. Indeed, any such vertex is at distance at most~$3$ from any other vertex in $S\setminus S'$, while every vertex in $S\setminus S'$ is at distance at least~$2$ from any vertex in $C$.
So, let us consider one vertex from $A^{\delta}$ for some $\delta\in \{\alpha, \beta, \gamma\}$ that lies in $S$, say, without loss of generality, $w\in A_i^{\alpha}$.
Note that $d(w,u'_i)=4$ and recall that $u'_i\in S$.
For any $q\in [m]$, there is a path of length $4$ between $w$ and $u'_i$ that covers $c_q$ if there is $v\in T^{\alpha}\cup F^{\alpha}$ such that both $(w,v)$ and $(v,c_q)$ are in $E(G)$.
But, by the construction, such an edge $(w,v)\in E(G)$ corresponds to an assignment of a variable that occurs in $c_q$, i.e., $(v,c_q)\in E(G)$ if the corresponding assignment to $v$ (\true\ or \false) satisfies the clause $C_q$ of the instance $\psi$.
Since $S$ is a geodetic set, for each $q\in[m]$, there is a vertex $w\in \bigcup_{\delta\in \{\alpha, \beta, \gamma\}}A^{\delta} \cap S$ that covers $c_q$ by a shortest path to some $u\in U'$.
Thus, let $\pi: X^{\alpha}\cup X^{\beta} \cup X^{\gamma} \rightarrow \{\true, \false\}$ be the retrieved assignment from the partial assignments that correspond to such vertices $w$ that are in both $S$ and $\bigcup_{\delta\in \{\alpha, \beta, \gamma\}}A^{\delta}$, and that is completed by selecting an arbitrary assignment for the variables in the buckets $X_i^{\delta}$ where $A_{i}^{\alpha}\cap S=\emptyset$.

Finally, as we observed above, for each $q\in[m]$, $c_q$ is covered, and thus, the constructed assignment $\pi$ satisfies all the clauses in $\calC$.
\end{proof}

\begin{lemma}\label{lem:GS-vc-backward}
If $\psi$ is a satisfiable \textsc{$3$-Partitioned-$3$-SAT} formula,  then
$G$ admits a geodetic set of size $k$.
\end{lemma}
\begin{proof}
Suppose $\pi: X^{\alpha}\cup X^{\beta} \cup X^{\gamma} \rightarrow \{\true, \false\}$ is a satisfying assignment for $\psi$.
We construct a geodetic set $S$ of size $k$ for $G$ using this assignment. 
Initially, let $$S=\{u_i'\mid i\in [\sqrt{n}]\}\cup \{b^{\delta}_{i,1}, b^{\delta}_{i,2}\mid \delta\in \{\alpha, \beta, \gamma\}, i\in [\sqrt{n}]\}.$$
At this point, $|S|=7\sqrt{n}$. 
Now, for each $i\in [\sqrt{n}]$ and $\delta\in \{\alpha, \beta, \gamma\}$, we add one vertex from $A^{\delta}_i$ to $S$ in the following way. 
For the bucket of variables $X^{\delta}_i$, consider how the variables of $X^{\delta}_i$ are assigned by $\pi$, and denote this assignment restricted to $X^{\delta}_i$ by $\pi^{\delta}_i$. 
Since $A^{\delta}_i$ contains a vertex for each of the possible $2^{\sqrt{n}}$ assignments and each of those corresponds to a certain assignment of $X^{\delta}_i$, we will find $w\in A^{\delta}_i$ that matches the assignment $\pi^{\delta}_i$. 
Then, we include this $w$ in $S$ as well. 
At the end, $|S|=10\sqrt{n}$.

Now, we show that $S$ is indeed a geodetic set of $G$.
First, recall that the vertices of $S$ are covered by any shortest path between them and any another vertex in $S$.
Further, recall that the neighbors of the degree-$1$ vertices in $S$ are also covered by the shortest paths between their degree-$1$ neighbor in $S$ and any another vertex in $S$. 
In the following case analysis, we omit the cases just described above. In each case, we consider sets of vertices that we want to cover by shortest paths between pairs of vertices in $S$. 

\begin{itemize}
\item {\bf Case 1:} $\boldsymbol{A_i^{\delta}\cup \{g_i^{\delta}\} \mid i\in [\sqrt{n}], \delta\in \{\alpha, \beta, \gamma\}\}}$\textbf{.}
For each $i\in [\sqrt{n}]$ and $\delta\in \{\alpha, \beta, \gamma\}$, there is a shortest path of length~$4$ between $b_{i, 1}^{\delta}$ and $b_{i, 2}^{\delta}$ that contains $g_i^{\delta}$ (any vertex in~$A_i^{\delta}$, respectively).

\item {\bf Case 2:} $\boldsymbol{T^{\delta}\cup F^{\delta}}$ \textbf{for any} $\boldsymbol{\delta\in \{\alpha, \beta, \gamma\}}$\textbf{.}
For each $\delta\in \{\alpha, \beta, \gamma\}$ and $i,i'\in [\sqrt{n}]$ such that $i\neq i'$, there is a shortest path of length~$6$ between $b_{i, 1}^{\delta}$ and $b_{i', 1}^{\delta}$ that is as follows.
First, it goes to $a_{i, 1}^{\delta}$ and then through $g^{\delta}_i$ to $w\in T^{\delta}\cup F^{\delta}$, and then through $g_{i'}^{\delta}$ to $a_{i', 1}^{\delta}$, before finishing at $b_{i', 1}^{\delta}$.
Since, for each $i\in [\sqrt{n}]$ and $\delta\in \{\alpha, \beta, \gamma\}$, $g_i^{\delta}$ is adjacent to all the vertices in $T^{\delta}\cup F^{\delta}$, the described path of length $6$ covers $T^{\delta}\cup F^{\delta}$.

\item {\bf Case 3:} $\boldsymbol{\{c_q\}_{q\in [m]}}$ \textbf{and} $\boldsymbol{V^{\delta}}$ \textbf{for any} $\boldsymbol{\delta\in \{\alpha, \beta, \gamma\}}$\textbf{.} 
For all $i\in [\sqrt{n}]$ and $\delta\in \{\alpha, \beta, \gamma\}$, consider the vertex $w\in A_i^{\delta}\cap S$. Recall that this $w$ corresponds to the assignment $\pi_i^{\delta}$, i.e., $\pi$ that is restricted to the subset of variables $X_i^{\delta}$.
First, there is a shortest path of length~$4$ between $w$ and $u'_i$ that contains $v_i^{\delta}$, $u_{i'}$ (for some $i'\in [\sqrt{n}]$ such that $i\neq i'$), and $u_i$.
Second, for each $q\in [m]$, there exists $i\in [\sqrt{n}]$ and $\delta\in \{\alpha, \beta, \gamma\}$ such that there is a shortest path of length $4$ from $w\in A_{i}^{\delta}$ to $u'_i$ that covers $c_q$.
Consider the variable that satisfied the clause $C_q$ of the initial instance $\psi$ under the assignment $\pi$, and, without loss of generality, let it be $x_{i, j}^{\delta}$.
Then, consider the bucket $X_i^{\delta}$ and select $w\in A_i^{\delta}$ such that $w$ corresponds to $\pi_i^{\delta}$.
Since $w$ corresponds to $\pi_i^{\delta}$, if $\pi(x^{\delta}_{i, j})=\true$ ($\pi(x^{\delta}_{i, j})=\false$, respectively), then $(w,t^{\delta}_j)\in E(G)$ ($(w,f^{\delta}_j)\in E(G)$, respectively) and, since $x_{i, j}^{\delta}=\true$ ($x_{i, j}^{\delta}=\false$, respectively) satisfies $C_q$, $(t^{\delta}_{j},c_q)\in E(G)$ ($(f^{\delta}_{j},c_q)\in E(G)$, respectively) as well.
Thus, we have a shortest path of length~$4$ that goes from $w$ to $t^{\delta}_j$ ($f^{\delta}_j$ in the latter case) to $c_q$ to $u_i$ to $u_i'$.
This way, all the vertices in $\{c_q\mid q\in [m]\}$ are also covered.
\end{itemize}

This covers all the vertices in $V(G)$, and thus, $S$ is a geodetic set of $G$.
\end{proof}

\begin{proof}[Proof of \Cref{thm:reduction-3-SAT-GS-VC}.] 
In \cref{subsec:reduction-3-Par-3-SAT-Geod-Set}, we presented a reduction that takes an instance $\psi$ of \textsc{$3$-Partitioned-$3$-SAT} and returns an equivalent instance $(G,k)$ of \textsc{Geodetic Set} (by Lemmas~\ref{lem:GS-vc-forward} and \ref{lem:GS-vc-backward}) in $2^{\OO(\sqrt{n})}$ time, where $k=10\sqrt{N}$. 
Note that $V(G)=2^{\OO(\sqrt{n})}$. 
Further, note that taking all the vertices in $B^{\delta}$, $V^{\delta}$, $T^{\delta}$, $F^{\delta}$, $U$, $C$, and $g^{\delta}_i$ for all $i\in [\sqrt{n}]$ and $\delta \in \{\alpha, \beta, \gamma\}$, results in a vertex cover of $G$. Hence, 

$$\vc(G) \leq 3 \cdot (|B^{\alpha}| + |V^{\alpha}| + |T^{\alpha}| + |F^{\alpha}| + \sqrt{n}) + |U| + |C|=\OO(\sqrt{n}).$$

Thus, $\vc(G)+k=\OO(\sqrt{n})$.
\end{proof}

\section{Algorithms for Vertex Cover Parameterization}
\label{sec:algo-vc-MD-GS-SMD}
%\section{\mdfull: Algorithms for Vertex Cover Parameterization}
\subsection{Algorithm for \mdfull}
\label{sec:algo-fpt-vertex-cover}

To prove Theorem~\ref{thm:algo-vertex-cover} for \mdfull, we first show that the following reduction rule is safe.

\begin{reduction rule} 
	\label{reduc:twins}
	If there exist three vertices $u,v,  x\in I$ such that $u,v, x$ are false twins, 
	then delete $x$ and decrease $k$ by one.
\end{reduction rule}
\begin{proof}[Proof that Reduction Rule~\ref{reduc:twins} is safe.]
	Since $u,v, x$ are false twins, $N(u)=N(v)=N(x)$. 
	This implies that, for any vertex $w\in V(G)\setminus \{u,v, x\}$, $d(w,v)=d(w,u)=d(w, x)$. 
	Hence, any resolving set that excludes at least two vertices in $\{u, v, x\}$
	cannot resolve all three pairs $\{u,v\}$,  $\{u, x\}$,  and $\{v, x\}$.
	As the vertices in $\{u, v, x\}$ are distance-wise indistinguishable
	from the remaining vertices, we can assume, without loss of generality, that 
	any resolving set contains both $u$ and $x$.
	Hence, any pair of vertices in $V(G) \setminus \{u, x\}$ that 
	is resolved by $x$ is also resolved by $u$.
	In other words, if $S$ is a resolving set of $G$, then
	$S \setminus \{x\}$ is a resolving set of $G - \{x\}$.
	This implies the correctness of the forward direction.
	The correctness of the reverse direction trivially follows from the fact
	that we can add $x$ into a resolving set of $G - \{x\}$
	to obtain a resolving set of $G$.
\end{proof}

\begin{lemma}
\label{lemma:kernel-vc}
\textsc{Metric Dimension},  parameterized by the vertex cover number $\vc$,
admits a polynomial-time kernelization algorithm that returns an instance
with 
$2^{\OO(\vc)}$ vertices.
\end{lemma}
\begin{proof}
Given a graph $G$, let $X\subseteq V(G)$ be a minimum vertex cover of $G$. 
If such a vertex cover is not given, then we can find a $2$-factor approximate vertex
cover in polynomial time. 
Let $I:=V(G)\setminus X$.
By the definition of a vertex cover, the vertices of $I$ are pairwise non-adjacent.

The kernelization algorithm exhaustively applies Reduction Rule~\ref{reduc:twins}.
Now, consider an instance on which~\cref{reduc:twins} is not applicable.
If the budget is negative, then the algorithm returns a trivial \no-instance
of constant size.
Otherwise, for any $Y\subseteq X$, there are at most two vertices $u,v\in I$ such that $N(u)=N(v)=Y$.
This implies that the number of vertices in the reduced instance
is at most $|X| + 2 \cdot 2^{|X|} =  2^{\vc +1} + \vc$.
\end{proof}

Next, we present an \XP-algorithm parameterized by the vertex 
cover number. 
This algorithm, along with the kernelization algorithm above,
imply\footnote{Note that the application of~\cref{reduc:twins} does not increase the vertex cover number.} that \textsc{Metric Dimension} admits
an algorithm running in $2^{\calO(\vc^2)}\cdot n^{\calO(1)}$ time.

\begin{lemma}
\label{lemma:algo-xp-vc}
\textsc{Metric Dimension} admits an algorithm running in $n^{\calO(\vc)}$ time.
\end{lemma}
\begin{proof}
The algorithm starts by computing a minimum vertex cover $X$ of $G$ 
in $2^{\mathcal{O}(\vc)} \cdot n^{\calO(1)}$ time using an \FPT\ algorithm
for the \textsc{Vertex Cover} problem, for example the one in~\cite{DBLP:journals/jal/ChenKJ01} or~\cite{HarrisN24}.
Let $I := V(G)\setminus X$. 
Then,  in polynomial time, it computes a largest subset $F$ of $I$ such that, 
for every vertex $u$ in $F$,  $I \setminus F$ contains a false twin of $u$.
By the arguments in the previous proof,  if there are false twins in $I$,  say $u, v$,
then any resolving set contains at least one of them.
Hence,  it is safe to assume that any resolving set contains $F$.
%If $k' \ge |X|$,  then the algorithm returns \yes.
If $k - |F| < 0$, then the algorithm returns \no.
Otherwise,  it enumerates every subset of vertices of size at most $|X|$ in $X \cup (I \setminus F)$. 
If there exists a subset $A \subseteq X \cup (I \setminus F)$ 
such that $A\cup F$ is a resolving set of $G$ of size at most $k$,
then it returns $A\cup F$.
Otherwise,  it returns \no.

In order to prove that the algorithm is correct, we prove that $X \cup F$ is a resolving set of $G$.
It is easy to see that, for a pair of distinct vertices $u,v$,
if $u \in X \cup F$ and $v \in V(G)$,  then the pair is resolved by $u$.
It remains to argue that every pair of distinct vertices in $(I \setminus F) \times (I \setminus F)$ is
resolved by $X \cup F$.
Note that, for any two vertices $u, v\in I \setminus F$,  $N(u) \neq N(v)$ as otherwise 
$u$ can be moved to $F$, contradicting the maximality of $F$.
Hence, there is a vertex in $X$ that is adjacent to $u$, but not adjacent to $v$,
resolving the pair $\langle u, v\rangle$.
This implies the correctness of the algorithm.
The running time of the algorithm easily follows from its description. 
\end{proof}
%\section{\gsfull: Algorithms for Vertex Cover Parameterization}
\subsection{Algorithm for \gsfull}
\label{sec:algo-fpt-vertex-cover-GS}

To prove Theorem~\ref{thm:algo-vertex-cover} for \gsfull, we start with the following fact about false twins.

%We start with the following fact about \emph{simplicial vertices} (a vertex is simplicial if its neighborhood forms a clique) and \emph{false twins} (vertices are false twins if they share the same open neighborhood). 
%%
%% \begin{observation}\label{obs:simplicial}
%% If a graph $G$ contains a simplicial vertex $v$, then $v$ belongs to any geodetic set of $G$.
%% \end{observation}
%% \begin{proof}
%% Observe that $v$ does not belong to any shortest path between any pair $x,y$ of vertices (both distinct from $v$).
%% \end{proof}

\begin{lemma}\label{lemma:open-twins}
If a graph $G$ contains a set $T$ of false twins that are not true twins and not simplicial, then any minimum-size geodetic set contains at most four vertices of $T$.
\end{lemma}
\begin{proof}
  Let $T=\{t_1,\ldots,t_h\}$ be a set of false twins in a graph $G$, that are not true twins and not simplicial. Thus, $T$ forms an independent set, and there are two non-adjacent vertices $x,y$ in the neighborhood of the vertices in $T$. Toward a contradiction, assume that $h\geq 5$ and $G$ has a minimum-size geodetic set $S$ that contains at least five vertices of $T$; without loss of generality, assume $\{t_1,\ldots, t_5\}\subseteq S$. We claim that $S'=(S\setminus\{t_1,t_2,t_3\})\cup\{x,y\}$ is still a geodetic set, contradicting the choice of $S$ as a minimum-size geodetic set of $G$.

  To see this, notice that any vertex from $V(G)\setminus T$ that is covered by some pair of vertices in $T\cap S$ is also covered by $t_4$ and $t_5$. Similarly, any vertex from $V(G)\setminus T$ covered by some pair $\langle t_i,z \rangle$ in $(S\cap T)\times (S\setminus T)$, is still covered by $t_4$ and $z$. Moreover, $x$ and $y$ cover all vertices of $T$, since they are at distance~2 from each other and all vertices in $T$ are their common neighbors. Thus, $S'$ is a geodetic set, as claimed.  
\end{proof}

\begin{lemma}
\label{lemma:kernel-vc-GS}
\gsfull, parameterized by the vertex cover number $\vc$, admits a polynomial-time kernelization algorithm that returns an instance with $2^{\OO(\vc)}$ vertices.
\end{lemma}
\begin{proof}
Given a graph $G$,  let $X\subseteq V(G)$ be a minimum-size vertex cover of $G$. 
If this vertex cover is not given, then we can find a $2$-factor approximate vertex cover in polynomial time.
Let $I:=V(G)\setminus X$; $I$ forms an independent set. The kernelization algorithm exhaustively applies the following reduction rules in a sequential manner.

\begin{reduction rule} 
\label{reduc:simplicial}
If there exist three simplicial vertices in $G$ that are false twins or true twins, then delete one of them from $G$ and decrease $k$ by one.
\end{reduction rule}

\begin{reduction rule} 
\label{reduc:open-twins}
If there exist six vertices in $G$ that are false twins but are not true twins nor simplicial, then delete one of them from $G$.
%If there exist five vertices in $G$ that are false twins but not simplicial, then delete one of them from $G$.
%\ftodo{can also be proved with 5 instead of 6, but proof becomes slightly longer}
\end{reduction rule}

To see that Reduction Rule~\ref{reduc:simplicial} is correct, assume that $G$ contains three simplicial vertices $u,v,w$ that are twins (false or true). We show that $G$ has a geodetic set of size $k$ if and only if the reduced graph $G'$, obtained from $G$ by deleting $u$, has a geodetic set of size $k-1$. For the forward direction, let $S$ be a geodetic set of $G$ of size $k$. By Observation~\ref{obs:simplicial}, $S$ contains each of $u,v,w$. Now, let $S'=S\setminus\{u\}$. This set of size $k-1$ is a geodetic set of $G'$. Indeed, any vertex of $G'$ that was covered in $G$ by $u$ and some other vertex $z$ of $S$, is also covered by $v$ and $z$ in $G'$. Conversely, if $G'$ has a geodetic set $S''$ of size $k-1$, then it is clear that $S''\cup\{u\}$ is a geodetic set of size $k$ in $G$.

For Reduction Rule~\ref{reduc:open-twins}, assume that $G$ contains six false twins (that are not true twins nor simplicial) as the set $T=\{t_1,\ldots,t_6\}$, and let $G'$ be the reduced graph obtained from $G$ by deleting $t_1$. We show that $G$ has a geodetic set of size $k$ if and only if $G'$ has a geodetic set of size $k$. For the forward direction, let $S$ be a minimum-size geodetic set of size (at most) $k$ of $G$. By Lemma~\ref{lemma:open-twins}, $S$ contains at most four vertices from $T$; without loss of generality, $t_1$ and $t_2$ do not belong to $S$. Since the distances among all pairs of vertices in $G'$ are the same as in $G$, $S$ is still a geodetic set of $G'$. Conversely, let $S'$ be a minimum-size geodetic set of $G'$ of size (at most) $k$. Again, by Lemma~\ref{lemma:open-twins}, we may assume that one vertex among $t_2,\ldots,t_6$ is not in $S'$, say, without loss of generality, that it is $t_2$. Note that $S'$ covers (in $G$) all vertices of $G'$. Thus, $t_2$ is covered by two vertices $x,y$ of $S'$. But then, $t_1$ is also covered by $x$ and $y$, since we can replace $t_2$ by $t_1$ in any shortest path between $x$ and $y$. Hence, $S'$ is also a geodetic set of $G$. 

%FF: PROOF FOR 5 INSTEAD OF 6
%For Reduction Rule~\ref{reduc:open-twins}, assume that $G$ contains five false twins as the set $T=\{t_1,\ldots,t_5\}$, and let $G'$ be the reduced graph obtained from $G$ by deleting $t_1$. We show that $G$ has a geodetic set of size $k$ if and only if $G'$ has a geodetic set of size $k$. Let $S$ be a geodetic set of size $k$ of $G$. By Lemma~\ref{lemma:open-twins}, $S$ contains at most four vertices among $T$; without loss of generality, $t_1$ does not belong to $S$. Since the distances among all pairs in $G'$ are the same as in $G$, $S$ is still a geodetic set of $G'$. Conversely, let $S$ be a geodetic set of $G'$ of size $k$. Similarly, $S'$ covers (in $G$) all vertices of $G'$. If $t_1$ is also covered, we are done. So, assume next that $t_1$ is not covered. If any other vertex in $\{t_2,\ldots,t_5\}$ is not in $G$ by $S'$, then it would also not be covered by $S'$ in $G'$, a contradiction. So, all of $t_2,\ldots,t_5$ are in $S'$. As in the proof of Lemma~\ref{lemma:open-twins}, let $x,y$ be two non-adjacent vertices in the neighborhood of $t_1$ (they exist since $t_1$ is not simplicial), and let $S''=S'\setminus\{t_2,t_3\}\cup\{x,y\}$. Notice that all vertices not in $T$ that were covered by one of $t_2,t_3$ and some $z$ are still covered by $t_4$ and $z$, and all vertices of $T$ are now covered by $x,y$. Thus, $S''$ is a geodetic set of size $k$ in $G$.

Now, consider an instance on which the reduction rules cannot be applied. If $k<0$, then we return a trivial \no-instance (for example, a single-vertex graph). Otherwise, notice that any set of false twins in $I$ contains at most five vertices. Hence, $G$ has at most $|X| + 5 \cdot 2^{|X|} = 2^{\OO(\vc)}$ vertices.
\end{proof}

Next, we present an \XP-algorithm parameterized by the vertex cover number. Together with Lemma~\ref{lemma:kernel-vc-GS}, they imply Theorem~\ref{thm:algo-vertex-cover} for \gsfull.

\begin{lemma}
\label{lemma:algo-xp-vc-GS}
\gsfull admits an algorithm running in $n^{\calO(\vc)}$ time.
\end{lemma}
\begin{proof}
The algorithm starts by computing a minimum vertex cover $X$ of $G$ in $2^{\mathcal{O}(\vc)} \cdot n^{\calO(1)}$ time using an \FPT\ algorithm for the \textsc{Vertex Cover} problem, for example the one in~\cite{DBLP:journals/jal/ChenKJ01} or~\cite{HarrisN24}. Let $I := V(G)\setminus X$.

In polynomial time, we compute the set $S$ of simplicial vertices of $G$. By Observation~\ref{obs:simplicial}, any geodetic set of $G$ contains all simplicial vertices of $G$. %Thus, if $k-|S|<0$, we return \no.
Now, notice that $X\cup S$ is a geodetic set of $G$. Indeed, any vertex $v$ from $I$ that is not simplicial has two non-adjacent neighbors $x,y$ in $X$, and thus, $v$ is covered by $x$ and $y$ (which are at distance~2 from each other).

Hence, to enumerate all possible minimum-size geodetic sets, it suffices to enumerate all subsets $S'$ of vertices of size at most $|X|$ in $(X \cup I) \setminus S$, and check whether $S\cup S'$ is a geodetic set. If one such set is indeed a geodetic set and has size at most $k$, we return \yes. Otherwise, we return \no. The statement follows.
\end{proof}

\section{Conclusion}\label{sec:conclu}

%perhaps other metric problems can be mentioned (I thought of scattered set but I think it's already covered by https://arxiv.org/abs/1709.02180 with a 2^vc algorithm). There is a tight 2^{vc log vc} algorithm for Steiner Forest, also accepted at ICALP (https://arxiv.org/abs/2402.09835).

We have seen that both \mdfull and \gsfull have a non-trivial
$2^{\Theta(\vc^2)}$ running-time dependency (unless the \ETH\ fails) in the vertex cover number
parameterization. Both problems are \FPT\ for related parameters, such
as vertex integrity, treedepth, distance to (co-)cluster, distance to
cograph, etc., as more generally, they are \FPT\ for cliquewidth plus
diameter~\cite{GHK22,KK22}. For both problems, it was proved that the
correct dependency in treedepth (and treewidth plus diameter) is in fact
double-exponential~\cite{DBLP:journals/corr/abs-2307-08149}, a fact
that is also true for feedback vertex set plus diameter for
\mdfull~\cite{DBLP:journals/corr/abs-2307-08149}. For distance to
(co-)cluster, algorithms with double-exponential dependency were given
for \mdfull in~\cite{GKIST23}. For the parameter max leaf number
$\ell$, the algorithm for \mdfull from~\cite{E15} uses ILPs, with a
dependency of the form $2^{\calO(\ell^6\log \ell)}$ (a similar
algorithm for \gsfull with dependency $2^{\calO(f\log f)}$ exists for
the feedback edge set number $f$~\cite{KK22}), which is unknown to be
tight. What is the correct dependency for all these parameters?
In particular, it seems interesting to determine for which parameter(s) the
jump from double-exponential to single-exponential dependency occurs.

For the related problem \textsc{Strong Metric Dimension}, the correct
dependency in the vertex cover number is known to be
double-exponential~\cite{DBLP:journals/corr/abs-2307-08149}. It would
be nice to determine whether similarly intriguing behaviors can be
exhibited for related metric-based problems, such as \textsc{Strong
  Geodetic Set}, whose parameterized complexity was recently adressed
in~\cite{DFPT22,lima2022computational}. Perhaps our techniques are
applicable to such related problems.

\bibliography{bib}

\begin{thebibliography}{10}

\bibitem{DBLP:journals/toct/AgrawalLSZ19}
A.~Agrawal, D.~Lokshtanov, S.~Saurabh, and M.~Zehavi.
\newblock Split contraction: The untold story.
\newblock {\em {ACM} Trans. Comput. Theory}, 11(3):18:1--18:22, 2019.

\bibitem{BelmonteFGR17}
R.~Belmonte, F.~V. Fomin, P.~A. Golovach, and M.~S. Ramanujan.
\newblock Metric dimension of bounded tree-length graphs.
\newblock {\em {SIAM} J. Discrete Math.}, 31(2):1217--1243, 2017.

\bibitem{BDM23}
B.~Bergougnoux, O.~Defrain, and F.~{Mc~Inerney}.
\newblock Enumerating minimal solution sets for metric graph problems.
\newblock In {\em Proc. of the 50th International Workshop on Graph-Theoretic
  Concepts in Computer Science ({WG} 2024)}, volume 14760 of {\em Lecture Notes
  in Computer Science}, pages 50--64. Springer, 2024.

\bibitem{BP21}
E.~Bonnet and N.~Purohit.
\newblock Metric dimension parameterized by treewidth.
\newblock {\em Algorithmica}, 83:2606--2633, 2021.

\bibitem{BDP23}
N.~Bousquet, Q.~Deschamps, and A.~Parreau.
\newblock Metric dimension parameterized by treewidth in chordal graphs.
\newblock In {\em Proc. of the 49th International Workshop on Graph-Theoretic
  Concepts in Computer Science ({WG} 2023)}, volume 14093 of {\em Lecture Notes
  in Computer Science}, pages 130--142. Springer, 2023.

\bibitem{floISAAC20}
D.~Chakraborty, S.~Das, F.~Foucaud, H.~Gahlawat, D.~Lajou, and B.~Roy.
\newblock {Algorithms and complexity for geodetic sets on planar and chordal
  graphs}.
\newblock In {\em Proc. of the 31st International Symposium on Algorithms and
  Computation (ISAAC 2020)}, volume 181 of {\em LIPIcs}, pages 7:1--7:15.
  Schloss Dagstuhl--Leibniz-Zentrum f{\"u}r Informatik, 2020.

\bibitem{DBLP:journals/corr/abs-2402-08346}
D.~Chakraborty, F.~Foucaud, D.~Majumdar, and P.~Tale.
\newblock Tight (double) exponential bounds for identification problems:
  Locating-dominating set and test cover.
\newblock In {\em Proc. of the 35th International Symposium on Algorithms and
  Computation (ISAAC 2024)}, volume 322 of {\em LIPIcs}, pages 19:1--19:18.
  Schloss Dagstuhl--Leibniz-Zentrum f{\"u}r Informatik, 2024.

\bibitem{CCMR23}
J.~Chalopin, V.~Chepoi, F.~{Mc~Inerney}, and S.~Ratel.
\newblock Non-clashing teaching maps for balls in graphs.
\newblock In {\em Proc. of the 37th Annual Conference on Learning Theory
  ({COLT} 2024)}, volume 247 of {\em Proceedings of Machine Learning Research},
  pages 840--875. {PMLR}, 2024.

\bibitem{DBLP:conf/iwpec/ChandranIK16}
L.~S. Chandran, D.~Issac, and A.~Karrenbauer.
\newblock On the parameterized complexity of biclique cover and partition.
\newblock In {\em Proc. of the 11th International Symposium on Parameterized
  and Exact Computation ({IPEC} 2016)}, volume~63 of {\em LIPIcs}, pages
  11:1--11:13. Schloss Dagstuhl - Leibniz-Zentrum f{\"{u}}r Informatik, 2016.

\bibitem{CHZ02}
G.~Chartrand, F.~Harary, and P.~Zhang.
\newblock On the geodetic number of a graph.
\newblock {\em Networks}, 39(1):1--6, 2002.

\bibitem{DBLP:journals/jal/ChenKJ01}
J.~Chen, I.~A. Kanj, and W.~Jia.
\newblock Vertex cover: Further observations and further improvements.
\newblock {\em J. Algorithms}, 41(2):280--301, 2001.

\bibitem{DBLP:books/sp/CyganFKLMPPS15}
M.~Cygan, F.~V. Fomin, L.~Kowalik, D.~Lokshtanov, D.~Marx, M.~Pilipczuk,
  M.~Pilipczuk, and S.~Saurabh.
\newblock {\em Parameterized Algorithms}.
\newblock Springer, 2015.

\bibitem{DBLP:journals/siamcomp/CyganPP16}
M.~Cygan, M.~Pilipczuk, and M.~Pilipczuk.
\newblock Known algorithms for edge clique cover are probably optimal.
\newblock {\em {SIAM} J. Comput.}, 45(1):67--83, 2016.

\bibitem{D12}
R.~Diestel.
\newblock {\em Graph Theory, 6th Edition}, volume 173 of {\em Graduate texts in
  mathematics}.
\newblock Springer, 2024.

\bibitem{dourado2010}
M.~C. Dourado, F.~Protti, D.~Rautenbach, and J.~L. Szwarcfiter.
\newblock Some remarks on the geodetic number of a graph.
\newblock {\em Discrete Mathematics}, 310(4):832--837, 2010.

\bibitem{DFPT22}
M.~Dumas, F.~Foucaud, A.~Perez, and I.~Todinca.
\newblock On graphs coverable by \({k}\) shortest paths.
\newblock {\em SIAM J. Discrete Math.}, 38(2):1840--1862, 2024.

\bibitem{E15}
D.~Eppstein.
\newblock Metric dimension parameterized by max leaf number.
\newblock {\em Journal of Graph Algorithms and Applications}, 19(1):313--323,
  2015.

\bibitem{ELW15}
L.~Epstein, A.~Levin, and G.~J. Woeginger.
\newblock The (weighted) metric dimension of graphs: Hard and easy cases.
\newblock {\em Algorithmica}, 72(4):1130--1171, 2015.

\bibitem{DBLP:journals/corr/abs-2307-08149}
F.~Foucaud, E.~Galby, L.~Khazaliya, S.~Li, F.~{Mc~Inerney}, R.~Sharma, and
  P.~Tale.
\newblock Problems in {NP} can admit double-exponential lower bounds when
  parameterized by treewidth or vertex cover.
\newblock In {\em Proc. of the 51st International Colloquium on Automata,
  Languages, and Programming ({ICALP} 2024)}, volume 297 of {\em LIPIcs}, pages
  66:1--66:19. Schloss Dagstuhl - Leibniz-Zentrum f{\"{u}}r Informatik, 2024.

\bibitem{GKIST23}
E.~Galby, L.~Khazaliya, F.~{Mc~Inerney}, R.~Sharma, and P.~Tale.
\newblock Metric dimension parameterized by feedback vertex set and other
  structural parameters.
\newblock {\em SIAM J. Discrete Math.}, 37(4):2241--2264, 2023.

\bibitem{GJ79}
M.~R. Garey and D.~S. Johnson.
\newblock {\em Computers and Intractability - A guide to NP-completeness}.
\newblock W.H. Freeman and Company, 1979.

\bibitem{GHK22}
T.~Gima, T.~Hanaka, M.~Kiyomi, Y.~Kobayashi, and Y.~Otachi.
\newblock Exploring the gap between treedepth and vertex cover through vertex
  integrity.
\newblock {\em Theoretical Computer Science}, 918:60--76, 2022.

\bibitem{DBLP:journals/tcs/GutinRRW20}
G.~Z. Gutin, M.~S. Ramanujan, F.~Reidl, and M.~Wahlstr{\"{o}}m.
\newblock Alternative parameterizations of metric dimension.
\newblock {\em Theoretical Computer Science}, 806:133--143, 2020.

\bibitem{harary1993}
F.~Harary, E.~Loukakis, and C.~Tsouros.
\newblock The geodetic number of a graph.
\newblock {\em Mathematical and Computer Modelling}, 17(11):89--95, 1993.

\bibitem{HM76}
F.~Harary and R.~A. Melter.
\newblock On the metric dimension of a graph.
\newblock {\em Ars Combinatoria}, 2:191--195, 1976.

\bibitem{HarrisN24}
D.~G. Harris and N.~S. Narayanaswamy.
\newblock A faster algorithm for vertex cover parameterized by solution size.
\newblock In {\em Proc. of the 41st International Symposium on Theoretical
  Aspects of Computer Science ({STACS} 2024)}, volume 289 of {\em LIPIcs},
  pages 40:1--40:18. Schloss Dagstuhl - Leibniz-Zentrum f{\"{u}}r Informatik,
  2024.

\bibitem{HartungN13}
S.~Hartung and A.~Nichterlein.
\newblock On the parameterized and approximation hardness of metric dimension.
\newblock In {\em Proc. of the 28th Conference on Computational Complexity,
  {CCC~2013}}, pages 266--276. {IEEE} Computer Society, 2013.

\bibitem{JKST19}
L.~Jaffke, O.-J. Kwon, T.~J.~F. Strømme, and J.~A. Telle.
\newblock {Mim-width III. Graph powers and generalized distance domination
  problems}.
\newblock {\em Theoretical Computer Science}, 796:216--236, 2019.

\bibitem{KLP22}
I.~Katsikarelis, M.~Lampis, and V.~Th. Paschos.
\newblock Structurally parameterized $d$-scattered set.
\newblock {\em Discrete Applied Mathematics}, 308:168--186, 2022.

\bibitem{KK22}
{L.} {Kellerhals} and {T.} {Koana}.
\newblock Parameterized complexity of geodetic set.
\newblock {\em Journal of Graph Algorithms and Applications}, 26(4):401--419,
  2022.

\bibitem{DBLP:journals/talg/KratschPR16}
S.~Kratsch, G.~Philip, and S.~Ray.
\newblock Point line cover: The easy kernel is essentially tight.
\newblock {\em {ACM} Trans. Algorithms}, 12(3):40:1--40:16, 2016.

\bibitem{DBLP:journals/corr/abs-2302-09604}
M.~Lampis, N.~Melissinos, and M.~Vasilakis.
\newblock Parameterized max min feedback vertex set.
\newblock In {\em Proc. of the 48th International Symposium on Mathematical
  Foundations of Computer Science ({MFCS} 2023)}, volume 272 of {\em LIPIcs},
  pages 62:1--62:15. Schloss Dagstuhl - Leibniz-Zentrum f{\"{u}}r Informatik,
  2023.

\bibitem{LM21}
S.~Li and M.~Pilipczuk.
\newblock Hardness of metric dimension in graphs of constant treewidth.
\newblock {\em Algorithmica}, 84(11):3110--3155, 2022.

\bibitem{lima2022computational}
C.~V. G.~C. Lima, V.~F. dos Santos, J.~H.~G. Sousa, and S.~A. Urrutia.
\newblock On the computational complexity of the strong geodetic recognition
  problem.
\newblock {\em RAIRO Oper. Res.}, 58(5):3755--3770, 2024.

\bibitem{MPP18}
D.~Marx, M.~Pilipczuk, and M.~Pilipczuk.
\newblock On subexponential parameterized algorithms for steiner tree and
  directed subset {TSP} on planar graphs.
\newblock In {\em Proc. of the 59th {IEEE} Annual Symposium on Foundations of
  Computer Science ({FOCS} 2018)}, pages 474--484, 2018.

\bibitem{DBLP:conf/mfcs/Pilipczuk11}
M.~Pilipczuk.
\newblock Problems parameterized by treewidth tractable in single exponential
  time: {A} logical approach.
\newblock In {\em Proc. of the 36th International Symposium on Mathematical
  Foundations of Computer Science ({MFCS} 2011)}, volume 6907 of {\em Lecture
  Notes in Computer Science}, pages 520--531. Springer, 2011.

\bibitem{DBLP:journals/iandc/SauS21}
I.~Sau and U.~dos Santos~Souza.
\newblock Hitting forbidden induced subgraphs on bounded treewidth graphs.
\newblock {\em Inf. Comput.}, 281:104812, 2021.

\bibitem{Slater75}
P.~J. Slater.
\newblock Leaves of trees.
\newblock In {\em Proc. of the {S}ixth {S}outheastern {C}onference on
  {C}ombinatorics, {G}raph {T}heory, and {C}omputing}, pages 549--559.
  Congressus Numerantium, No. XIV. Utilitas Mathematica, 1975.

\bibitem{DBLP:journals/corr/abs-2403-03501}
P.~Tale.
\newblock Double exponential lower bound for telephone broadcast.
\newblock {\em CoRR}, abs/2403.03501, 2024.
\newblock URL: \url{https://doi.org/10.48550/arXiv.2403.03501}.

\end{thebibliography}

\end{document}